\documentclass[accepted=2025-03-18,a4paper]{quantumarticle}
\pdfoutput=1

\usepackage[german,english]{babel}
\usepackage[utf8]{inputenc}
\usepackage[colorinlistoftodos, color=green!40, prependcaption]{todonotes}
\usepackage{braket}

\PassOptionsToPackage{compress}{natbib}
\usepackage[numbers]{natbib}
\usepackage{amsthm,amsfonts,amsmath,amssymb}
\usepackage{mathtools}
\usepackage{physics}
\usepackage{xcolor}
\usepackage{graphicx}
\usepackage[left=23mm,right=13mm,top=35mm,columnsep=15pt]{geometry} 
\usepackage{adjustbox}
\usepackage{placeins}
\usepackage[T1]{fontenc}
\usepackage{lipsum}
\usepackage{csquotes}
\usepackage{orcidlink}
\usepackage{soul}
\usepackage{bbold}
\usepackage{nicefrac}
\usepackage{floatrow}

\usepackage{varioref}
\labelformat{figure}{Fig.\;#1}
\DeclareUnicodeCharacter{2009}{~}

\newcommand{\ie} {i.e.~}
\newcommand{\eg} {e.g.~}
\newcommand{\Eqs} {Eqs.~}
\newcommand{\Eq} {Eq.~}
\newcommand{\cf} {cf.~}

\DeclareMathOperator{\hc}{H.c.}

\DeclareMathOperator{\sgn}{sgn}

\newcommand{\aref}[1]{\hyperref[#1]{Appendix~\ref*{#1}}}
\newcommand{\eq}{Eq.\,}
\newcommand{\eqs}{Eqs.\,}

\newcommand{\rref}{Ref.\,}

\begin{document}
\title{Dipole-dipole interactions mediated by a photonic flat band}

\author{Enrico Di Benedetto\,\orcidlink{0009-0003-3613-5257}}
    \email{enrico.dibenedetto@unipa.it}
    \affiliation{Universit$\grave{a}$  degli Studi di Palermo, Dipartimento di Fisica e Chimica-Emilio Segrè, Via Archirafi 36, 90123 Palermo (Italy)}

\author{Alejandro González-Tudela\,\orcidlink{0000-0003-2307-6967}}
    \email[Correspondence email address: ]{a.gonzalez.tudela@csic.es}
    \affiliation{CSIC – Instituto de Fisica Fundamental, C. de Serrano 113, 28006 Madrid (Spain)}

\author{Francesco Ciccarello\,\orcidlink{0000-0002-6061-1255}}
    \email[Correspondence email address: ]{francesco.ciccarello@unipa.it}
    \affiliation{Universit$\grave{a}$  degli Studi di Palermo, Dipartimento di Fisica e Chimica-Emilio Segrè, Via Archirafi 36, 90123 Palermo (Italy)}
    \affiliation{NEST, Istituto Nanoscienze-CNR, Piazza S. Silvestro 12, 56127 Pisa (Italy)}

\date{} 

\begin{abstract}
Flat bands (FBs) are energy bands with zero group velocity, which in electronic systems were shown to favor strongly correlated phenomena. Indeed, a FB can be spanned with a basis of strictly localized states, the so called \textit{compact localized states} (CLSs), which are yet generally non-orthogonal.
Here, we study emergent dipole-dipole interactions between emitters dispersively coupled to the photonic analogue of a FB, a setup within reach in state-of the-art experimental platforms. We show that the strength of such photon-mediated interactions decays exponentially with distance with a characteristic localization length  which, unlike typical behaviours with standard bands, saturates to a finite value as the emitter's energy approaches the FB. Remarkably, we find that the localization length grows with the overlap between CLSs according to an analytically-derived universal scaling law valid for a large class of FBs both in 1D and 2D. Using giant atoms (non-local atom-field coupling) allows to tailor interaction potentials having the same shape of a CLS or a superposition of a few of these.
\end{abstract}

\maketitle

\section{Introduction} 
\label{sec:introduction}

The coherent interaction between quantum emitters and engineered low-dimensional photonic environments is a hot research area of modern quantum optics, especially in the emerging framework of waveguide QED \cite{roy_colloquium_2017,chang_colloquium_2018,sheremet_waveguide_2023,ciccarello_waveguide_2024,gonzalez-tudela_lightmatter_2024}. A core motivation is unveiling qualitatively new paradigms of atom-photon interaction, well beyond those occurring in standard electromagnetic environments (like free space or low-loss cavities). These can potentially be exploited to implement cutting-edge quantum information processing tasks or to favor observation of new quantum many-body phenomena. This relies on the fact that photonic baths with tailored properties and dimensionality, \eg photonic versions of electronic tight-binding models, can today be implemented in a variety of experimental scenarios. 
This includes photonic crystals in the optical domain \cite{bouscal_systematic_2024,goban_atomlight_2014,zhou_coupling_2023,menon_integrated_2023}, superconducting circuits in the microwaves \cite{jouanny_band_2024,liu_quantum_2017,mirhosseini_superconducting_2018,scigliuzzo_controlling_2022} and matter-wave emulators \cite{lanuza_multiband_2022,krinner_spontaneous_2018}, allowing the study even of lattices with non-trivial properties such as occurrence of topological phases (even in 2D) \cite{kim_quantum_2021,owens_chiral_2022} or able to emulate curved spaces \cite{kollar_hyperbolic_2019}.

A key phenomenon to appreciate the effect of an engineered photonic bath are photon-mediated dispersive interactions.
For instance, while a high-finesse cavity can mediate an effective all-to-all interaction between atoms in the dispersive regime \cite{ritsch_cold_2013,zheng_efficient_2000}, replacing the cavity with an engineered {\it lattice} of coupled cavities can result in interactions with a non-trivial shape of the interatomic potential when the atoms are tuned in a photonic bandgap \cite{douglas_quantum_2015,gonzalez-tudela_subwavelength_2015}. The resulting interaction range can be controlled via modulation the detuning from the band edge, which has been experimentally confirmed in circuit QED \cite{sundaresan_interacting_2019,kim_quantum_2021,scigliuzzo_controlling_2022,zhang_superconducting_2023} and predicted to be a resource for efficient quantum simulations \cite{armon2021,bello2022,tabares2022} and implementation of hybrid quantum-classical algorithms \cite{tabares_variational_2023}.
Such effective coherent interactions between atoms can be understood as being mediated by atom-photon bound states (BSs) formed by a single quantum emitter  \cite{lambropoulos_fundamental_2000,lombardo_photon_2014,calajo_atom-field_2016,shi_bound_2016,gonzalez-tudela_markovian_2017,liu_quantum_2017,gonzalez-tudela_exotic_2018,gonzalez-tudela_non-markovian_2018,gonzalez-tudela_anisotropic_2019,krinner_spontaneous_2018,sanchez-burillo_single_2019,roman-roche_bound_2020}, whose corresponding photonic wavefunction (typically exponentially localized around the atom) in fact shapes the spatial profile of the effective interatomic potential \cite{douglas_quantum_2015,shi_effective_2018,leonforte_quantum_2024}. Depending on the bath structure, this can exhibit an unconventional or even exotic dependence on the emitters' positions which can be accompanied by topological protection  \cite{Bello2019,LeonfortePRL2021,VegaPRA21,roccati2022exotic,bello2,roccati2024hermitian,tian2024power}.

Remarkably, due to destructive interference mechanisms, some lattices can host a special type of bands having in fact the same spectrum as a (one-mode) cavity. Such a band is called {\it flat band} (FB) in that its dispersion law is flat. Accordingly, a FB has zero width so that its spectrum features only one frequency just like a perfect cavity mode.
Such kind of bands are well-known to show up in certain natural and artificial lattice structures  \cite{vicencio_poblete_photonic_2021,leykam_artificial_2018}, a prominent example occurring in the celebrated quantum Hall effect \cite{klitzing_new_1980}.
FBs are currently investigated in condensed matter and photonics \cite{danieli_flat_2024} because, due to their macroscopically large degeneracy and effective quenching of kinetic energy, they can favor emergence of many-body effects, highly-correlated phases of condensed matter \cite{hu_correlated_2023} and non-linearity \cite{rivas_seltrapping_2020}. Moreover, FBs are extremely sensitive to disorder \cite{leykam_flat_2013}.
Typically, a FB arises when the lattice structure is such that, for each unit cell, one can construct a {\it compact} stationary state (\ie strictly localized on a few sites), which exactly decouples from the rest of the lattice by destructive interference. As a hallmark, these states, called \textit{Compact Localised States} (CLSs) \cite{rhim_classification_2019,chen_impossibility_2014,read_compactly_2017} in most cases are non-orthogonal and can be used to construct a basis of localized states spanning the FB space that is alternative to the canonical basis of Bloch states (these being in contrast orthogonal and unbound).

Given the above framework, here we tackle the question as to what kind of photon-mediated interactions are expected when atoms are dispersively coupled to a photonic FB, and in particular whether they are like those in the vicinity of a standard band (\ie with finite width) or instead like typical ones in cavity QED. The question is non-trivial as a FB has the same dimensionality of a standard band but, in contrast, zero bandwidth. On the other hand, a FB has a spectrum comprising only one frequency like a one-mode cavity but contains a thermodynamically large number of modes.

With these motivations, this paper presents a general study of atom-photon interactions in the presence of photonic FBs \footnote{Atoms coupled to photonic FBs in specific models appeared in recent studies
\cite{de_bernardis_light-matter_2021, bienias_circuit_2022, de_bernardis_chiral_2023}.}, by focusing in particular on photon-mediated interactions whose associated potential shape is inherited (as for standard bands) from the shape of atom-photon BSs. In contrast to the edge of a standard band, it turns out that in the vicinity of a FB the shape of the BS is insensitive to the detuning, while the interaction range of photon-mediated interactions generally remains finite even when the atom energetically approaches the FB.

Importantly, we connect the atom-photon BS to the form of the photonic CLSs characteristic of the considered band by demonstrating that the overlap between CLSs provides the mechanism which enables atoms in different cells to mutually interact. We derive an analytical and exact general relationship between the BS localization length and the non-orthogonality of CLSs valid for a large class of 1D and 2D lattices.
Our theory describes occurrence of atom-photon bound states and dipole-dipole photon-mediated interactions in the vicinity of a FB in the most general case, including when CLSs lack orthogonality.

This work is organized as follows. In Section \ref{model-sec}, we introduce the model and notation and review some basic concepts concerning atom-photon bound states and photon-mediated interactions. In Section \ref{sec-cs}, we present a case study illustrating some peculiar features of BSs in the presence of FBs: this anticipates some of the main original results of this work and at the same time allows the reader to first familiarize with the physics of flatbands. 
In Section \ref{sec-CLS}, we review the concept of Compact Localised States and introduce some 1D models used in this paper. 
This is then used in the following Sections \ref{sec:dressed-BS} and \ref{sec-touching} in order to discuss bound states seeded by a FB, which represents a central result of our investigation. Finally, the case of a giant atom coupled to a FB system and the ensuing photon-mediated interactions is addressed in Section mediated interactions is addressed in Section \ref{sec:giants}. Finally, in Section \ref{sec-conc} we draw our conclusions.

\section{Setup and review of photon-mediated interactions} 
\label{model-sec}

Here, we introduce the general formalism we will work with and review some basic notions and important examples of photon-mediated interactions. We start from atom-photon BSs, which are necessary in order to appreciate the physics in the presence of FBs to be presented later.

\subsection{General model}

We consider a quantum emitter modeled as a two-level system whose pseudo-spin ladder operator is $\sigma = \ketbra{g}{e}$, with $\ket{g}$ and $\ket{e}$ being respectively the ground and excited states, whose energy difference is $\omega_{0}$. The atom is locally coupled under the rotating-wave approximation to a photonic bath modeled as a set of single-mode coupled cavities or resonators each labeled by index $x$ (which is generally intended as a set of indexes). Accordingly, the total Hamiltonian reads
\begin{equation}
    \label{eq:tot_lm}
    H = \omega_{0} \sigma^\dagger  \sigma +  H_B + g \left(a^\dagger_{x_0}\sigma + \hc \right),
\end{equation}
where $x_0$ labels the cavity which the emitter is coupled to with strength $g$, assumed to \textit{weak} in a sense that will be specified better in the following sections.
The free Hamiltonian of the photonic bath $H_B$ reads 
\begin{align}
    \label{eq:CCA_mult}
    H_B = \sum_{x} \omega_{x} a^\dagger_{x} a_{x} 
    + \sum_{x\not=x'} J_{xx'}\qty(a^\dagger_{x'} a_{x} + \hc ),
\end{align}
where $\omega_{x}$ is the bare frequency of the $x$th cavity, $a_{x}$ ($a^\dagger_x$) the associated creation (destruction) bosonic ladder operator while $J_{xx'}$ denotes the photon hopping rate between cavities $x$ and $x'$. The Fock state where cavity $x$ has one photon (with all the remaining ones having zero photons) will be denoted as $\ket{x} = a^\dagger_x \ket{\rm vac}$, where $\ket{\rm vac}$ is the vacuum state of the field.

While many properties discussed in this paper require solely that the bath $B$ possesses a FB, which can happen even if the photonic bath is not  translationally-invariant, all of the examples that we will discuss concern photonic lattices. In these cases, $B$ is a $D$-dimensional lattice so that index $x$ in the above equations should be intended as the pair of indexes $(n,\nu)$, where $n$ stands for $D$ integers labeling a primitive unit cells around the Bravais lattice vector $\bf{r}_n$, while $\nu$ labels the sublattices. All lengths have to be intended in units of the lattice constant.
The normal frequencies of $B$ (in the thermodynamic limit) comprise a series of {\it bands}, labeled by index $m$ and with corresponding dispersion law $\omega_{m}(k)$, where the $D$-dimensional wave vector $k$ lies in the first Brillouin zone (to make notation lighter, we will write $k$ in place of ${\bf k}$ whenever possible). 
Accordingly, the bath Hamiltonian can be written in diagonal form as
\begin{equation}
    \label{eq:Hdiag}
    H_B=\sum_{k,m} \omega_{m}(k) \Psi_{k,m}^{\dagger}\Psi_{k,m}\,,
\end{equation}
where $\Psi_{k,m}$ and $\Psi^\dagger_{k,m}$ are ladder operators associated with the $m$th band.

Throughout this work, we will focus on the single-excitation sector, \ie the subspace spanned by the set $\left\{ \ket{e}\ket{\rm vac}\!, \ket{g} \ket{x} \right\}$ (with $x$ running over all cavities).
For the sake of simplicity, we will adopt a lighter notation in what follows and replace
\begin{align*}
    \ket{e}\ket{{\rm vac}}\rightarrow \ket{e}, && \ket{g} \ket{x}\rightarrow \ket{x}.
\end{align*}
Hence, from now on $\ket{e}$ will be the state where the excitation lies on the atom and the field has no photons, while $\ket{x}$ is the state where the atom in the ground state $\ket{g}$ and a single photon at cavity $x$ has one photon (with each of the remaining cavities in the vacuum state). In particular, $\ket{x_0}$ is the state where a single photon lies at cavity $x_0$ (the one directly coupled to the emitter).

\subsection{Atom-photon bound states}
\label{bs-sec}

Within the single-excitation sector, an atom-photon bound state (BS) $\ket{\Psi_{\rm BS}}$ is a normalized dressed state such that $H\ket{\Psi_{\rm BS}}=\omega_{\rm BS}\ket{\Psi_{\rm BS}}$, where $\omega_{\rm BS}$ is a real solution of the pole equation
$$\omega_{\rm BS} = \omega_{0} + g^2 \mel{x_0}{G_B(\omega_{\rm BS})}{x_0},$$
while the wavefunction (up to a normalization factor) reads \cite{lambropoulos_fundamental_2000}
\begin{align}
	\label{psibs-1}
	\ket{\Psi_{\rm BS}} \propto  \ket{e}+ \ket{\psi_{\rm BS}} &&
	\text{with}&&
	\ket{\psi_{\rm BS}} \!= \!g\, G_B (\omega_{\rm BS})\!\ket{x_0}\!.
\end{align}
Here, $G_B(\omega)$ is the bath resolvent or Green's function in the single-excitation subspace \cite{lambropoulos_fundamental_2000}, whose general definition is [\cf\eq\eqref{eq:Hdiag}]
\begin{equation}
	\label{green}
	G_B(\omega)=\sum_{k,m} \frac{\dyad{\Psi_{k,m}}}{\omega-\omega_{m}(k)}.
\end{equation}
When $B$ is not a lattice, index $(k,m)$ is just replaced by the index(es) labeling the eigenmodes of the field.

Typically, a BS occurs when the atom is coupled off-resonantly to $B$ (as in all examples to be discussed in this paper where $\omega_{0}$ will be tuned within a bandgap of bath $B$). In this case, to leading order in the coupling strength we have 
\begin{align}
    \label{psibs}
    \omega_{\rm BS} = \omega_{0}\,, && 
    \ket{\psi_{\rm BS}} = g\, G_B (\omega_{0})\ket{x_0}.
\end{align}
Wavefunction $\ket{\psi_{\rm BS}}$ describes a single photon localized around the atom's location $x_0$.

The simplest, and in some respects trivial, example of an atom-photon BS occurs in a {\it single-mode cavity}, \ie when $H_B = \omega_c \,a_{x_0}^\dagger a_{x_0}$. In this case, the bath Green's function simply reads $G_B(\omega)=\dyad{x_0}/(\omega-\omega_c)$ so that the BS reduces to $\ket{\Psi_{\rm BS}} = \ket{e} + \frac{g}{\omega_{0}-\omega_c} \ket{x_0}$ \footnote{This is one of the two dressed states in the single-excitation sector, specifically the one with a dominant atomic component.  The other state instead has energy $\simeq \omega_{c}$ and is mostly photonic.}.

Another paradigmatic instance of BS occurs when $B$ is a 1D lattice of coupled cavities described by the Hamiltonian $H_B = J \sum_n a_{x_n}^\dag a_{x_{n+1}} + \hc$ with $J>0$. The energy spectrum (in the thermodynamic limit) consists of a single band in the interval $\comm{-2J}{2J}$ with dispersion law $\omega(k) = 2 J \cos k$. 
The corresponding Green's function (for $\abs{\omega}> 2J$, \ie out of band) reads \cite{economou_greens_1979,arfken_chapter_2013} 
\begin{equation}
    \label{GBa}
    \begin{split}
        \mel{x_n}{G_B(\omega)}{x_m} &= \frac{1}{N} \sum_k \frac{e^{ik(x_n-x_m)}}{\omega-2J \cos k} =\\ &\to \frac{(-1)^{|x_n-x_m|}}{2\sqrt{J\delta}}\exp(-\frac{|x_n-x_m|}{\lambda}),
    \end{split}
\end{equation}
where the last equality holds in the thermodynamical limit ($N\to \infty$) and for $\delta=\omega-2J>0$ (above the upper band edge) and $\delta\ll J $. From here, we see that the BS is exponentially localized with
\begin{equation}
    \label{BS-array}
    \begin{split}
        \lambda = \sqrt{\frac{J}{\delta}}.
    \end{split}
\end{equation}
Thus, the photon is exponentially localized around the atom over a region of characteristic length $\lambda$ (\textit{localization length}). Remarkably, this scales as $\sim \delta^{-\nicefrac{1}{2}}$, entailing that, as the atomic frequency approaches the band edge (\ie for $\delta\rightarrow 0^+$), the BS gets more and more delocalized.
Later on, we will see that in the presence of a FB a very different behavior occurs with the BS localization length saturating to a constant value as $\delta\rightarrow 0^+$.

\subsection{Effective Hamiltonian with many emitters}\label{sec-Heff}

In the presence of many identical emitters, each indexed by $j$, the Hamiltonian \eqref{eq:tot_lm} is naturally generalized as
\begin{equation}
    H=\omega_{0}\sum_j\sigma_j^\dag\sigma_j+H_B+g\sum_{j}  (a_{x_j} ^\dag \sigma_{j }+\hc)\label{H1-many},
\end{equation}
with $x_j$ labeling the cavity which the $j$th atom is coupled.

When $\omega_{0}$ lies within a photonic bandgap and $B$ is in the vacuum state, the field degrees of freedom can be adiabatically eliminated and the dynamics of the emitters is fully described by the effective many-body Hamiltonian \cite{douglas_quantum_2015,gonzalez-tudela_subwavelength_2015,leonforte_quantum_2024}
\begin{equation}
    \label{eq:Heff-em}
    H_{\rm eff} = \sum_{ij} \left(\mathcal{K}_{ij} \sigma_i^\dagger \sigma_j + \hc \right),
\end{equation}
where
\begin{equation}
	\label{eq:kij}
	\mathcal{K}_{ij} = g \braket{x_i}{\psi_{{\rm BS},j}}.
\end{equation}
Here, $\ket{\psi_{{\rm BS},j}}$ is the BS seeded by the $j$th atom, whose expression is analogous to \eq\eqref{psibs} with $x_j$ in place of $x_0$.
Thus, when atoms are dispersively coupled to the bath $B$, they undergo a coherent mutual interaction described by the effective interatomic potential $\mathcal{K}_{ij}$. Significantly, \eq\eqref{eq:kij} shows that the interaction strength between a pair of atoms is simply proportional to the BS seeded by one emitter (as if this were alone) on the cavity where the other emitter sits in. Thus, the BS wavefunction of a single emitter in fact embodies the spatial shape of the photon-mediated interaction potential in the presence of many emitters in the dispersive regime.

Based on the discussion in Section \ref{bs-sec}, for a set of atoms all coupled to a single cavity we get $\mathcal{K}_{ij} = g^2/(\omega_0-\omega_c)$ irrespective of $i$ and $j$, meaning that the emitters undergo an all-to-all interaction. 

Instead, owing to the exponential localization of BS \eqref{BS-array}, a set of atoms dispersively coupled to a homogeneous cavity array (see Section \ref{bs-sec}) undergo an effective short-range interaction described by
\begin{equation}
    \label{short}
    \mathcal{K}_{ij} =\frac{(-1)^{|x_i-x_j|}}{\sqrt{J\delta}}\exp(-\frac{|x_i-x_j|}{\lambda})\,,
\end{equation}
where the BS localization length $\lambda$ [see \eq\eqref{BS-array}] can now be seen as the characteristic interaction range depending on $\omega_0$.

\section{Bound state near a photonic flat band: case study}\label{sec-cs}

\begin{figure*}[t!]
    \centering
    \includegraphics[width=\textwidth]{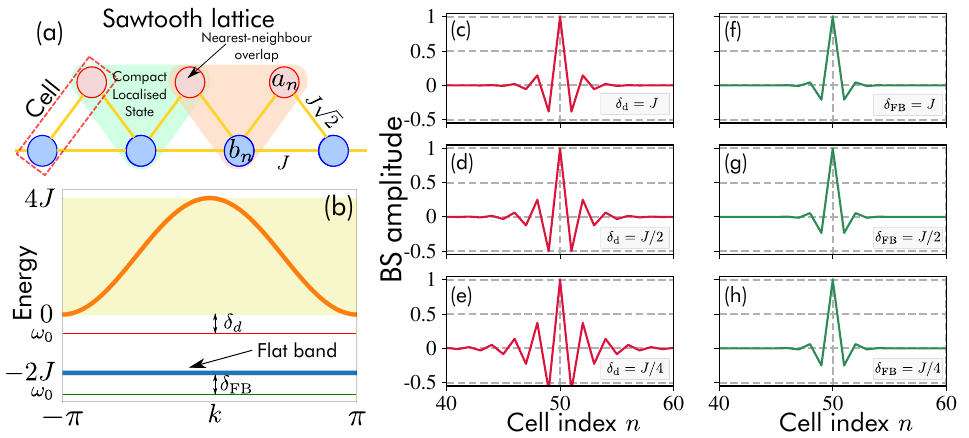}
    \caption{Atom-photon bound state in the photonic sawtooth lattice. (a)  Sketch of the sawtooth lattice, showing in particular a primitive unit cell (red dashed box) and the values of hopping rates between nearest-neighbour cavities (the bare frequency of each cavity is set to zero). Two overlapping CLSs are shown.
    (b) Frequency spectrum of the photonic lattice, comprising a dispersive band between $\omega=0$ and $\omega=4J$ and additionally a FB at $\omega_{\rm FB} = -2J$. We call $\delta_d$ ($\delta_{\rm FB}$) the detuning of the atomic frequency $\omega_{0}$ from the dispersive band (from the FB).
    (c)-(h) BS photonic wavefunction $|\psi_{\rm BS}\rangle$ [see \Eq \eqref{psibs-1}] for an atom coupled dispersively to the dispersive band [(c)-(e)] and to the FB [(f)-(h))] for decreasing values of $\delta_d$ ($\delta_{\rm FB}$). In each case, we plot only the amplitude on sublattice $a_n$ (the behaviour on sublattice $b_n$ is qualitatively similar) rescaled to the maximum value and considered $N=100$ unit cells under periodic boundary conditions with the atom coupled to cavity $a_{50}$ with strength $g=0.001J$.}
    \label{fig:saw}
\end{figure*}
A major difference between the two paradigmatic baths considered so far, \ie a cavity versus a coupled-cavity array, is that the latter can host {\it propagating} photons whose speed is proportional to the slope of the dispersion law $\omega(k)$. Notice that the presence of a non-trivial dispersion law $\omega(k)$ is essential for the occurrence of photon-mediated interactions with finite and $\omega_0$-dependent interaction range $\lambda$ [\cf\eqs\eqref{BS-array} and \eqref{short}]. 

In the remainder of this paper, we will deal with dispersive physics near a {\it flat band} (FB). A FB is a special photonic band whose dispersion law is $k$-independent, \ie 
\begin{equation}
    \label{wFB}
    \omega(k) = \rm{constant}.
\end{equation}
Thus, unlike a standard band, a FB has zero bandwidth, a feature shared with a (perfect) cavity. Differently from a cavity and analogously to a standard band, however, the FB energy has a thermodynamically large degeneracy (matching the number of lattice cells $N$) \footnote{If bath $B$ is not a lattice, a FB occurs when there exists a normal frequency with a thermodynamically large degeneracy. An instance of such a situation are certain kinds of hyperbolic systems, where there exists a FB gapped from the rest of the energy spectrum \cite{kollar_hyperbolic_2019}.}.

In order to introduce some typical properties of photon-mediated interactions around a FB, in this section we will consider the case study of the 1D sawtooth photonic lattice in \ref{fig:saw}(a) \cite{hyrkas_many-particle_2013,sanchez-burillo_chiral_2020}, arguably the simplest yet non-trivial system where a FB shows up. This is a bipartite lattice, where the $n$th cell consists of a pair of cavities labeled $a_n$ and $b_n$. Each cavity $b_n$ is coupled to its nearest neighbours $b_{n\pm1}$ with photon hopping rate $J>0$ and with rate $J\sqrt{2}$ to cavities $a_{n-1}$ and $a_n$
\footnote{Strictly speaking, this is a special case of the sawtooth lattice in which the ratio of the two hopping rates is constrained so as to ensure the emergence of a FB.}. All cavities have the same bare frequency which we set to zero.
The two bands of $B$ have dispersion laws (see \aref{subsec:sawtooth})
\begin{align}
    \omega_{\rm FB}= -2J, && \omega_{d}(k) = 2J (1+\cos k),\label{saw-w}
\end{align}
As shown in \ref{fig:saw}(b), the band labeled by subscript $d$ is a standard dispersive band of width 4$J$, below which a FB of energy $-2J$ stands out (the energy separation of this from the edge of band $d$ is $2J$).

We now discuss the atom-photon BS formed by an emitter dispersively coupled to the 1D sawtooth lattice in the two different regimes where the effect of either the FB or the dispersive band is negligible.
We start by tuning $\omega_0$ very far from the FB (so that this can be fully neglected) but relatively near the lower edge of the dispersive band so as to fulfill $g\ll \delta_d \ll \delta_{\rm FB}$ with $\delta_d$  and $\delta_{\rm FB}$ respectively the detuning from the lower edge of band $d$ and from the FB [see \ref{fig:saw}(b)].
The spatial profile of the BS photonic wavefunction is plotted in \ref{fig:saw}(c)-(e) for decreasing values of $\delta_d$. Similarly to \eq\eqref{BS-array} (homogeneous array), the BS is exponentially localized around the atom with a localization length which diverges as $\delta_d\rightarrow 0^+$. 

We next consider the regime $g\ll  \delta_{\rm FB}\ll \delta_d $ [see \ref{fig:saw}(b)] in a way that the atom is significantly (although dispersively) coupled only to the FB with the contribution of the dispersive band now negligible. As shown in \ref{fig:saw}(f)-(h), instead of diverging, the BS localization length now {\it saturates} to a finite value as $\delta_{\rm FB}\rightarrow 0^+$. Correspondingly, the BS wavefunction no longer changes when $\delta_{\rm FB}$ is small enough, which shows that this asymptotic BS is insensitive to the atom frequency. 

The simple instance just discussed suggests that atom-photon BSs in the vicinity of a photonic FB, along with the ensued photon-mediated interactions, have a quite different nature compared to standard photonic bands. While the fact that the BS remains localized is somewhat reminiscent of the behavior in a standard one-mode cavity, it is natural to wonder what the BS localization length depends on. It will turn out that this depends on the way in which the emitter is coupled to the lattice as well as some intrinsic properties of the FB, in particular the orthogonality of the so called \textit{Compact Localized States} (CLSs), a key concept in FB theory. For this reason, the next section is fully devoted to a review of CLSs, whose main properties will be illustrated through variegated examples. This will provide us with the necessary theoretical basis to formulate general properties of BSs in Section \ref{sec:dressed-BS}, including more general types of FBs (\eg occurring in 2D lattices).

\section{Compact localized states}\label{sec-CLS}

As for any band, the eigenspace made up by all photonic states where a single photon occupies a given FB can be naturally spanned by the Bloch stationary states $\{\ket{\Psi_{k,{\rm FB}}}\}$ with $H_B \ket{\Psi_{k,{\rm FB}}}=\omega_{{\rm FB}}\ket{\Psi_{k,{\rm FB}}}$, where $k$ runs over the first Brillouin zone.
 
 Accordingly, the projector onto the FB eigenspace in the Bloch states basis $\{\ket{\Psi_{k,{\rm FB}}}\}$ reads
\begin{equation}
    \label{eq:PSIPFB}
    \mathcal{P}_{\rm FB} = \sum_{k} \dyad{\Psi_{k,{\rm FB}}}.
\end{equation}
By definition, when $\mathcal{P}_{\rm FB}$ is applied to a generic single photon state it returns its projection onto the FB eigenspace. Notice that, being  translationally invariant (see Bloch theorem \cite{bloch_ber_1929}), each basis state $\ket{\Psi_{k,{\rm FB}}}$ is necessarily {\it unbound}, \ie its wavefunction has support on the entire lattice. Yet, since photons in a FB have zero group velocity [\cf\eq\eqref{wFB}], it is natural to expect the FB eigenspace to admit an alternative basis of {\it localized} (\ie bound) states $\qty{\ket{\phi_{n}}}$, one for each unit cell indexed by $n$ (when $B$ is a lattice). Such a basis indeed exists and its elements are called \textit{Compact Localised States} (CLSs) \cite{rhim_classification_2019} (an exception occurs when the FB touches a dispersive band, a pathological case that we will address later on).
Note that a CLS is compact in the sense that, typically, it is strictly localized only on a finite, usually small, number of neighbouring unit cells. There exists a general way to express CLSs in the basis of Bloch states $\{\ket{\Psi_{k,{\rm FB}}}\}$ for a $D$-dimensional latttice, which reads \cite{rhim_classification_2019}
\begin{equation}
	\label{eq:cls-expansion}
	\ket{\phi_n} = \frac{1}{N^{D/2}} \sum_{k \in {\rm BZ}} \sqrt{f(k)}\, e^{-i \vb{k} \cdot \vb{r}_n} \ket{\Psi_{k,{\rm FB}}},
\end{equation}
with $N$ the number of cells and where $f(k)\ge 0$ is a suitable function (recall that $\vb{r}_n$ is a Bravais lattice identifying a unit cell). 

The idea behind CLSs is somewhat similar to standard Wannier states \cite{wannier_structure_1937} in that a CLS can be expressed as a suitable superposition of unbound Bloch states that yields a state localized around a lattice cell. Unlike a Wannier state, however, a CLS is itself an eigenstate of the bath Hamiltonian, \ie $H_B\ket{\phi_{n}}=\omega_{{\rm FB}}\ket{\phi_{n}}$, since the Bloch states entering the expansion all have the same energy $\omega_{\rm FB}$ (reflecting the high FB degeneracy).

\subsection{Class $U$ and overlap parameters $\alpha_d$}
\label{sec-class}

Note that, while there is a one-to-one correspondence between CLSs and unit cells, a CLS $\ket{\phi_{n}}$ is generally spread over $U\geq 1$ cells, where the positive integer $U$ is the so called {\it class} of the CLS set \cite{maimaiti_universal_2019}. In the special case $U=1$, each $\ket{\phi_n}$ is entirely localized within the $n$th cell so that the CLSs do not overlap in space and are thus orthogonal, \ie $ \braket{\phi_{n}}{\phi_{n'}}=\delta_{nn'}$ (an instance is the double-comb lattice of \ref{fig:lattices}). For $U\ge 2$, instead, CLSs necessarily overlap in space with one another, which remarkably causes them to be generally {\it non-orthogonal}. 

It is easy to understand that, for a 1D (generally multipartite) lattice, CLSs of class $U=2$ are such that each CLS overlaps only its two nearest-neighbor CLSs, being orthogonal to the all the remaining ones (an instance is the sawtooth lattice of \ref{fig:saw} which we will discuss shortly). In such a case, the overlap between a pair of CLSs thus reads
\begin{equation}
    \label{eq:alpha_1d}
    \braket{\phi_n}{\phi_{n'}} = 1 + \alpha (\delta_{n,n'+1}+\delta_{n,n'-1}),
\end{equation}
where $\alpha = \braket{\phi_n}{\phi_{n\pm1}}$ is an overlap parameter.

For 2D lattices, it is easier to think of class $U$ as a 2D vector of integer components, describing how many cells the CLS is spread over along either direction. For instance, if $U=(2,2)$ the CLS will overlap only its nearest-neighbors along each direction (an instance being the checkerboard lattice of \ref{fig:checkerboard} to be discussed later on). 
In this case, the overlap between a pair of CLSs takes the form 
\begin{equation}
    \begin{split}
        \braket{\phi_n}{\phi_{n'}} &= 1 + \alpha_x (\delta_{n_x,n'_x+1}+\delta_{n_x,n'_x-1}) \\
        &+ \alpha_y (\delta_{n_y,n'_y+1}+\delta_{n_y,n'_y-1}),
    \end{split}
\end{equation}
which now features {\it two} overlap parameters, $\alpha_x$ and $\alpha_y$, corresponding to the $\vu{e}_x$ and $\vu{e}_y$ directions, respectively.
Likewise, for a generic $D$-dimensional model with $U = (2,2,\dots,2)$ (i.e., only nearest-neighbor overlaps between CLSs along \textit{any} $\vu{e}_d$ direction), the overlap reads
\begin{equation}
    \label{eq:def_alpha}
    \braket{\phi_n}{\phi_{n'}} = 1 + \sum_{d=1}^D \alpha_d (\delta_{n,n'+\vu{e}_d}+\delta_{n,n'-\vu{e}_d}),
\end{equation}
with $\alpha_d$ the overlap along the $d$th direction. Thus, in general, there are $D$ (generally) different overlap parameters $\alpha_d$, as many as the spatial dimensions of the lattice. A more general expression for CLSs of class $U>2$ can be found in \aref{app:proj}, \Eq \eqref{app:eq:Mij}.

Notice that, owing to the degeneracy of the FB energy $\omega_{\rm FB}$, the set of CLSs spanning the FB eigenspace for a given lattice is not unique. We call \textit{minimal} the set of CLSs having the lowest class $U$. In the remainder, we will only consider minimal CLSs.

\subsection{Instances of CLSs in 1D lattices}

In order to make the reader familiarize with CLSs and their properties (especially non-orthogonality), we present next some examples of lattices exhibiting FBs [see \ref{fig:lattices} and \ref{fig:1dkagome}]. In this section, we will only discuss 1D models, meaning that here the cell index $n$ is an integer and the wavevector $k$ a real number (see \aref{app:lattices} for more details on those models).

\subsubsection{Double-comb lattice}
\label{sec-db}

\begin{figure}[t]
    \centering
    \includegraphics[width=\textwidth]{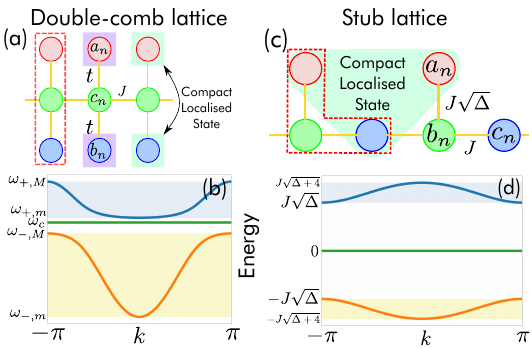}
    \caption{Instances of one-dimensional lattices with FB. (a) Double-comb lattice (the frequency of cavity $c_n$ is zero, while the one of cavities $a_n$ and $b_n$ is $\omega_c$). (b) Field spectrum of the double-comb lattice showing a FB at energy $\omega=\omega_c$. (c) Stub lattice (the frequency of all cavities is set to zero). (d) Field spectrum of the stub lattice, showing a zero-energy FB. In (a) and (c) we highlight the primitive unit cell (red dashed line) and possible choices of CLSs. Notice that CLSs are not overlapping in the double-comb lattice in panel (a), but they do overlap in the stub lattice (c).}
    \label{fig:lattices}
\end{figure}

The double-comb lattice is a tripartite lattice, which comprises three sublattices $a ,b$ and $c$ [see \ref{fig:lattices}(a)]. 
Each cavity $c_n$ of the central sublattice is coupled to the nearest-neighbour cavities $c_{n\pm1}$ with photon hopping rate $J$ and with rate $t$ to cavities $a_n$ and $b_n$ (upper and lower sublattices, respectively). 
Cavities $a_n$ and $b_n$ have the same bare frequency $\omega_c$, while the one of $c_n$ is set to zero. The spectrum of $B$, which is plotted in a representative case in \ref{fig:lattices}(b), features a FB at energy $\omega_{ \rm FB} = \omega_c$ and two dispersive bands. The occurrence of such FB is easy to predict since the antisymmetric state 
\begin{equation}
    \ket{\phi_n} = \frac{1}{\sqrt{2}} \big( \ket{a_n} - \ket{b_n}\big)
\end{equation} 
clearly decouples from state $\ket{c_n}$ (hence the rest of the lattice) due to destructive interference and is thereby an eigenstate of the bath Hamiltonian $H_B$ with energy $\omega_c$. 
Evidently, there exists one such state for each cell $n$, explaining the origin of the FB at energy $\omega_{ \rm FB} = \omega_c$. As each $\ket{\phi_n}$ is strictly localized within a unit cell [see \ref{fig:lattices}(a)], states $\{\ket{\phi_n}\}$ form a set of orthogonal CLSs of class $U=1$ based on the previous definition.

\subsubsection{Sawtooth lattice}
\label{sec-saw}

We already introduced the sawtooth lattice in Section \ref{sec-cs} and \ref{fig:saw}.
Somewhat similarly to the double-comb lattice, the FB at $\omega_{\rm FB} = - 2J$ also arises through destructive interference. Indeed, the superposition
\begin{equation}
    \label{eq:CLSsaw}
	\ket{\phi_n} = \frac{1}{2} \Big( \ket{a_n} + \ket{a_{n-1}} -\sqrt{2}\ket{b_n}\Big)
\end{equation}
decouples from sites $b_{n\pm1}$, hence the rest of the lattice.
Since we can build up one such state for each cell, the set $\{ \ket{\phi_n}\}$ form a basis spanning the FB eigenspace. Unlike the double-comb lattice, however, it is clear that each state $\ket{\phi_n}$ is not localized within a single unit cell and is {\it not} orthogonal to the two CLSs $\ket{\phi_{n\pm1}}$ [\cf\ref{fig:saw}(a)]. 
Indeed, it is easy to verify that \eq \eqref{eq:alpha_1d} holds in this case with $\alpha = 1/4$, meaning that this set is of class $U=2$.
Therefore, these CLSs form a \textit{non-orthogonal} basis of the FB eigenspace.

\subsubsection{Stub lattice}
\label{sec-stub}

The stub lattice (or 1D Lieb lattice) \cite{hyrkas_many-particle_2013,baboux_bosonic_2016,real_flat-band_2017} is the tripartite lattice sketched in \ref{fig:lattices}(c), where each cavity $b_n$ is coupled to cavities $c_{n\pm1}$ with rate $J$ and side-coupled to cavity $a_n$ with rate $J\sqrt{\Delta}$ where $\Delta\ge 0$ is a dimensionless parameter. The spectrum of $H_B$ is symmetric around $\omega=0$, at which energy a FB arises ($\omega_{\rm FB } = 0$). The gap separating the FB from each dispersive band is proportional to $J \sqrt{\Delta}$ [see \ref{fig:lattices}(c)].

Similarly to the previous lattices and CLSs, the origin of the zero-energy FB can be understood by noting that state
\begin{equation}
    \label{eq:CLSstub}
    \ket{\phi_n} = \tfrac{1}{\sqrt{2+\Delta}}\left( \ket{a_n} +  \ket{a_{n+1}} - \sqrt{\Delta} \ket{c_n}\right)
\end{equation}
decouples from the rest of the chain.
Like in the sawtooth lattice, the non-orthogonal set $\{\ket{\phi_n} \}$ form a CLS of class $U=2$
since \Eq \eqref{eq:alpha_1d} holds also in the present case with
\begin{equation}
    \label{eq:alphadelta}
    \alpha = \frac{1}{2+\Delta}.
\end{equation}
Notice that $\Delta$ controls both the overlap between CLSs and the energy gap between the FB and each dispersive band (such a tunability is not possible in the sawtooth model). For $\Delta\to0^+$ we get $\alpha \to 1/2 ^-$ and zero band gap. For growing $\Delta$, the non-orthogonality parameter $\alpha$ gets smaller and smaller [indeed $\ket{\phi_n}$ is more and more localized around cavity $c_n$, \cf\eqref{eq:CLSstub}] while the gap gets larger and larger.

Note that the stub lattice enjoys chiral symmetry \cite{ramachandran_chiral_2017} which guarantees a zero-energy FB to exist \cite{lieb_two_1989} even in the presence of disorder provided that it does not break chiral symmetry \footnote{This zero-energy FB always exists in lattices with chiral symmetry and an odd number of sublattices (Lieb's Theorem \cite{lieb_two_1989}). This is because chiral energy imposes the energy spectrum to be symmetric around $\omega(k)=0$ for any $k$. As the total number of energy bands is odd, one of the bands must necessarily be a zero-energy FB.}.

\begin{figure*}[t]
    \centering
    \includegraphics[width=0.95\textwidth]{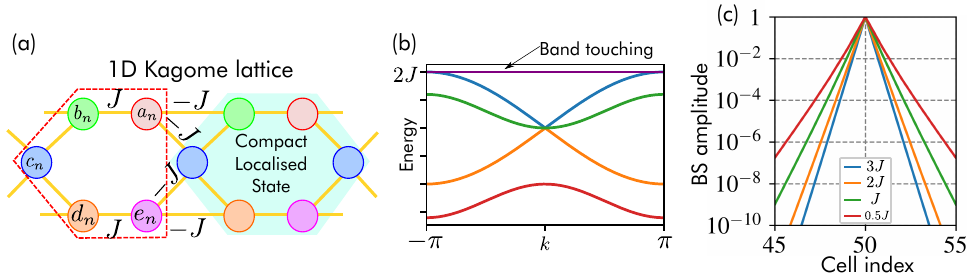}
    \caption{(a) Structure of the 1D Kagomè model, featuring five  sublattices with hopping rates $\pm J$. A possible choice of CLS is higlighted in green. (b) Energy spectrum featuring four dispersive bands and a FB of energy $\omega_{\rm FB}=2J$. The FB touches the upper edge of one of the dispersive bands. (c) Bound state seeded by an atom in the dispersive regime $\delta_{\rm FB}\gg g$ with $\delta_{\rm FB}=\omega_0-2J$ the detuning from the FB, which we report in the legend (we show only the BS wavefunction on sublattice $a$). We consider $100$ cells with the emitter coupled to cavity $a_{50}$.  In contrast \eg to the case in \ref{fig:saw} (f)-(h), the localization length here does depend on $\delta_{\rm FB}$ (which is due to the band touching as explained later on in Section \ref{sec-touching}).}
    \label{fig:1dkagome}
\end{figure*}

\subsubsection{1D Kagomè lattice}
\label{subsubsec:kag}

The 1D Kagomè model sketched in \ref{fig:1dkagome}(a) is a lattice with five sublattices, representing the 1D version of the popular 2D Kagomè model \cite{mekata_kagome_2003}. In this model each cavity is coupled to its nearest neighbour with rate $-J$, except for pairs ($a_n, b_n$) and ($d_n,e_n$) which are coupled with rate $J$. 
This system differs from the previous instances in that there exists a FB (of frequency $2J$) which touches the edge of a dispersive band [see \ref{fig:1dkagome}(b)]. 

Even in this lattice, one can construct CLSs of class $U=2$ having the form [see \ref{fig:1dkagome}(a)]
\begin{equation}
    \ket{\phi_n} = \frac{1}{\sqrt{6}} \Big[\ket{c_n} + \ket{c_{n+1}} - \ket{a_n} - \ket{b_n} - \ket{d_n} - \ket{e_n}\Big]\,,
\end{equation}
which fulfill \eq\eqref{eq:alpha_1d} with $\alpha=\braket{\phi_n}{\phi_{n\pm1}} = 1/6$.

\section{Bound states in a FB: general properties}
\label{sec:dressed-BS}

Armed with the notion of CLSs, we are now ready to establish general properties of atom-photon BSs. This and the following two sections contain the main results of our work.

\subsection{Atom-photon bound state as a superposition of compact localized states}

We recall that the BS shape is generally dictated by the field's bare Green's function [see \eq\eqref{psibs-1}]. 
As in the examples discussed in Section \ref{sec-cs}, here we consider the regime where the emitter is dispersively coupled to one specific band (not necessarily a FB), which we will call $\tilde{m}$, in such a way that the effects of all the other bands can be neglected. This in particular happens when, as in Section \ref{sec-cs}, there is a finite energy gap between band $\tilde{m}$ and all the other bands (no band crossing/touching). Another circumstance where this regime holds (as we will see later on) is when the contribution of band $\tilde{m}$ to the density of states (at energies close to $\omega_0$) is dominant compared to that from the other bands.

In such regime, we can approximate the BS wavefunction as  [\cf\eqs\eqref{psibs-1}, \eqref{green} and \eqref{psibs}]
\begin{equation}
    \label{psibs-2-in}
    \psi_{\rm BS}(x) = g \mel{x}{G_B (\omega_{0})}{x_0} \simeq g \sum_k \frac{\langle x\ket{\Psi_{k,\tilde{m}}}\bra{\Psi_{k,\tilde{m}}}x_0\rangle} {\omega_{0}-\omega_{\tilde{m}}(k)}
\end{equation}
(notice that only the contribution of band $\tilde{m}$ is retained).

Wavefunction $\psi_{\rm BS}(x)$, including its shape and width, will generally depend on the emitter frequency $\omega_{0}$, hence on its detuning from the $\tilde{m}$th band edge.
However, if band $\tilde{m}$ is a FB, in \eq\eqref{psibs-2-in} we can replace $\omega_{\tilde{m}}(k)$ with $\omega_{\rm FB}$, which is $k$-independent, and obtain the BS wavefunction (here $x$ labels a generic lattice site, even beyond 1D)
\begin{equation}
    \label{psibs-2}
    \psi_{\rm BS}(x) \simeq  \frac{g} {\omega_{0}-\omega_{\rm FB}}\mel{x}{\mathcal{P}_{\rm FB}}{x_0},
\end{equation}
where we substituted the projector $\mathcal{P}_{\rm FB}$ on the FB eigenspace [see \Eq \eqref{eq:PSIPFB}].
Evidently, the shape and width of the BSs become independent of the detuning of the atom from the FB in agreement with the behaviour emerging from \ref{fig:saw}(f)-(h). Notice that this independence holds also for an atom dispersively coupled to a standard cavity in which case however the BS is trivial.
A typical case where \Eq \eqref{psibs-2} holds is when $g \ll \delta_{\rm FB} \ll \delta_d$, where $g$ is the coupling strength while $\delta_{\rm FB}$ and $\delta_d$ are respectively the emitter's detuning from the FB and the edge of the nearest dispersive band (see e.g. Section \ref{sec-cs}), where it is understood that there exists a finite gap separating the FB from the dispersive bands. Interestingly, \Eq \eqref{psibs-2} can hold even if this gap vanishes, e.g. when the FB touches the edge of a dispersive band, provided that the FB  contribution to the density of states dominates over those due to dispersive bands: we will see an example of such a case in Section \ref{sec-touching}.

Having assessed the independence of the detuning, we next characterize the structure and spatial range of the BS (which will then also be those of photon-mediated interactions, see Section \ref{sec-Heff}). For this aim, taking advantage of \eq\eqref{eq:cls-expansion}, it is convenient to re-express the FB projector \eqref{eq:PSIPFB} in terms of the CLSs basis as (see \aref{app:proj} for details)
\begin{equation}
	\label{eq:CLSPFB}
	\mathcal{P}_{\rm FB} =\sum_{n,n'} \xi_{n n'} \ketbra{\phi_{n'}}{\phi_{n}} 
\end{equation}
with
\begin{equation}\label{xinn}
    \xi_{nn'}= \frac{1}{N^D} \sum_k \frac{1}{f(\vb{k}) }\,e^{i\vb{k}\cdot(\vb{r}_{n}-\vb{r}_{n'})}\,.
\end{equation}
Notice that the scalar product between two CLSs depends on function $f(\bf{k})$ as [\cf\eq\eqref{eq:cls-expansion}] 
\begin{equation}
    \label{eq:braket}
    \braket{\phi_{n}}{\phi_{n'}} = \frac{1}{N^D} \sum_{\vb{k}} f(\vb{k}) e^{i\vb{k}\cdot(\vb{r}_{n}-\vb{r}_{n'})}\,.
\end{equation}

Remarkably, unlike the Bloch-states expansion of Eq.~\eqref{eq:PSIPFB}, the decomposition of Eq.~\eqref{eq:CLSPFB} in terms of CLSs is {\it non}-diagonal. This is a consequence of the CLSs' non-orthogonality discussed in the previous section. Indeed, for $\braket{\phi_{n}}{\phi_{n'}}=\delta_{nn'}$ (orthogonal CLSs) we have $f(\vb{k}) = 1$ and hence $\xi_{nn'}=\delta_{nn'}$ in a way that \Eq \eqref{eq:CLSPFB} reduces to a standard diagonal expansion in the CLS basis. On the other hand, provided that only nearest-neighbour CLSs are overlapping (as in all the examples of this work), $f(\vb{k})$ is generally given by
\begin{equation}
    \label{fk}
    f(\vb{k}) = 1 + 2\sum_{d=1}^D \alpha_d \cos k_d
\end{equation}
where $D$ is the lattice dimension, $k_d$ the $d$th component of wave vector $\vb{k}$ and $\alpha_d$ the overlap between a pair of nearest-neighbor CLSs lying along the $d$th direction (see Section \ref{sec-class}). Expression \eqref{fk} can be easily derived by comparing \eq\eqref{eq:def_alpha} and \eq \eqref{eq:braket}.
It is clear that since $f(\vb{k}) \geq 0$ we have $\sum_d |\alpha_d| \leq 1/2$.

Finally, plugging \Eq \eqref{eq:CLSPFB} into \eq\eqref{psibs-2} yields
\begin{equation}
    \label{psibs-3}
    \begin{split}
       \psi_{\rm BS}(x) &\simeq \frac{g} {\omega_{0}-\omega_{\rm FB}} \sum_{n} w_{n}(x_0)  \phi_{n}(x) \,,
    \end{split}
\end{equation}
with 
\begin{equation}\label{weight}
    w_{n}(x_0)=\sum_{n'}\xi_{nn'}\phi^*_{n'}(x_0)
\end{equation}
which expresses the BS wavefunction as a weighted superposition of CLSs $\phi_{n}(x) = \braket{x}{\phi_{n}}$ where the weight function $w_n(x_0)$ is defined by \eq \eqref{weight} and depends on the cavity $x_0$ to which the emitter is coupled to. In particular, the weight function $w_n(x_0)$ is proportional to $\phi^*_{n'}(x_0) = \braket{\phi_{n'}}{x_0}$, which is non-zero only if cavity $x_0$ overlaps the CLS $\ket{\phi_{n'}}$.
As a consequence, being the CLSs strictly compact, the weight function $w_n$ [\cf\eq \eqref{weight}] involves only a finite number of terms. 

Expansion \eqref{psibs-3} is a central result of this work. Notice that for non-overlapping CLSs (class $U=1$) we have $w_n=\phi^*_n(x_0)$ (since $\xi_{nn'}=\delta_{nn'}$ as we saw previously), which entails that in this case the BS just coincides the CLS overlapping the cavity (if any). Hence, for $U=1$, atoms sitting in different cells will just not interact. This situation is reminescent of cavity QED, where CLSs play the role of non-overlapping cavity modes: atoms in isolated cavities have no way to cross-talk. In this sense, FBs of class $U=1$ show a cavity-like behaviour. In contrast, for $U\ge 2$, CLSs do overlap each other in a way that now $\xi_{nn'}\neq \delta_{nn'}$ and thus the BS is a superposition of more than one CLS. We see that the interaction between atoms located in different cells is now possible and this is a consequence of the CLSs' overlap. This situation is quite different from standard cavity QED. In particular, the CLSs cannot be interpreted as overlapping cavity modes since, if so, these would in fact couple the cavities affecting their spectrum non-trivially (in contrast to the present FB).

\begin{figure}
	\centering
	\includegraphics[width=0.75\textwidth]{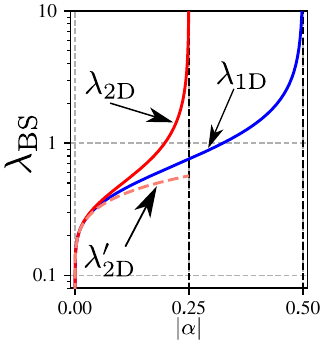}
        \caption{Universal scaling of the BS localization length $\lambda_{\rm BS}$ as a function of the CLSs overlap $|\alpha|$ for 1D and 2D lattices based on \eqs\eqref{eq:expLBS} and \eqref{eq:lbs2d}, respectively (in the common case that only nearest-neighbour CLSs overlap). Here, $|\alpha|=1/2$ and $|\alpha|=1/4$ are the maximum allowed values of $|\alpha|$, respectively in 1D and 2D, consistent with the hypothesis that only nearest-neighbour CLSs overlap. In 2D, the behaviour is the sum of two decaying exponentials with localization lengths $\lambda_{\rm 2D}$ and $\lambda'_{\rm 2D}$ [\cf\eq\eqref{eq:lbs2d}].}
	\label{fig:stub-FBBS}
\end{figure}
In the following, we consider the recurrent case where only nearest-neighbor CLSs overlap, which happens for $U=2$ for 1D lattices and $U=(2,2)$ for 2D ones (\cf Section \ref{sec-class}) and show that the BS is exponentially localized by deriving the localization length as an explicit general function of the CLSs' overlap. We point out however that the present theory is valid even when overlapping CLSs are not limited to nearest-neighbor ones, which is illustrated in \aref{app:u3} with an example of 1D tripartite lattice showing up a FB of class $U=3$.

\subsection{1D lattices, $U=2$}
\label{subsec:loclength}

In this case, Eq.~\eqref{fk} reduces to $f(k) = 1 +2\alpha \cos k$, with $|\alpha| = |\!\braket{\phi_n}{\phi_{n+1}}\!| \leq 1/2$. Plugging this into \eq\eqref{xinn} and working out the resulting integral in the thermodynamic limit ($N\to\infty$), one ends up with (see \aref{app:xi} for the proof)
\begin{equation}
    \label{eq:xi1d}
    \xi_{nn'}= \frac{(-\sgn{\alpha})^{|n-n'|}}{\sqrt{1-4\alpha^2}} \exp{-\frac{|n-n'|}{\lambda_{\rm 1D}}}, 
\end{equation}
where $\sgn{\alpha}$ denotes the sign of $\alpha$ while 
\begin{equation}
    \label{eq:expLBS}
   \frac{1}{\lambda_{\rm 1D}} = \mbox{settsech}(2|\alpha|),
\end{equation}
being $\mbox{settsech}(x)$ the inverse function of the hyperbolic secant. Therefore, $\xi_{nn'}$ decays exponentially with $|n-n'|$ with a characteristic length $\lambda_{\rm 1D}$ which is a growing function of the non-orthogonality coefficient $\alpha$ [see \ref{fig:stub-FBBS}(b) showing that $\lambda_{\rm 1D}$ vanishes for $\alpha=0$ and diverges for $|\alpha|\rightarrow 1/2$]. 
Based on a result derived in \aref{sec-corr}, we get that the BS has the same exponential shape and localization length as $\xi_{nn'}$. For instance, in the case of the sawtooth lattice we have $\alpha = 1/4$, which replaced in \eq\eqref{eq:xi1d} yields $\lambda_{\rm BS} \simeq 0.759$ in perfect agreement with the scaling observed in \ref{fig:saw}(f-h).

\subsection{2D lattices, $U=(2,2)$}
\label{sec:2d}

For a 2D lattice and a square geometry (we focus on this case as all our 2D examples have such structure), \Eq \eqref{fk} is given by $f(k_x,k_y) = 1 + 2\alpha_x\cos k_x + 2\alpha_y\cos k_y$. 
For the sake of argument, we focus on the isotropic case $\alpha_x = \alpha_y = \alpha$, (a generalization to $\alpha_x \neq \alpha_y$ is straightforward but expressions get involved).
One can then show that (see \aref{app:xi})
\begin{equation}\label{xinn2D}
    \xi_{nn'} = A \exp{-\frac{|\vb{r}_n-\vb{r}_{n'}|}{\lambda_{\rm 2D}}} + B \exp{-\frac{|\vb{r}_n-\vb{r}_{n'}|}{\lambda'_{\rm 2D}}},
\end{equation}
where $A,B$ are constants while
\begin{align}
    \label{eq:lbs2d}
    \frac{1}{\lambda_{\rm 2D}} = \mbox{settsech}\qty|\frac{2\alpha}{1-2\alpha}|,\,\,\,
    \frac{1}{\lambda'_{\rm 2D}} = \mbox{settsech}\qty|\frac{2\alpha}{1+2\alpha}| 
\end{align}
($\alpha$ is subject to the constraint $|\alpha|\leq 1/4$).
Thus, in this case the decay results from the superposition of two exponential functions with localization length $\lambda_{\rm 2D}$ and $\lambda'_{\rm 2D}$, respectively. As in 1D, both localization lengths grow with $|\alpha|$ (see \ref{fig:stub-FBBS}(b)]. Differently from 1D, however, when $\alpha$ approaches its maximum value 1/4 only $\lambda_{\rm 2D}$ diverges while $\lambda'_{\rm 2D}$ instead saturates to the finite value $\lambda'_{\rm 2D}(\alpha\rightarrow 1/4)\simeq 0.567$ (for $\alpha\rightarrow -1/4$, $\lambda_{\rm 2D}$ saturates to the same value while $\lambda'_{\rm 2D}$ diverges).

This means that in this limiting case the weight function $\xi_{nn'}$ is a single exponential having localization length equal to $\lambda\simeq 0.567$. In 2D, thereby, photon-mediated interactions are finite-ranged provided that the CLSs have non-zero overlap.

\section{Flatband touching a dispersive band}\label{sec-touching}

The conclusions developed so far apply to lattices featuring an energetically isolated FB, \ie separated by a finite gap from all the remaining bands [recall \eq\eqref{psibs-2} where only the contribution of the FB is retained]. This rules out in particular those lattices where a FB arises on the edge of a dispersive band.
In 1D, such band touching happens for example in the stub lattice for $\Delta = 0$ (see Section \ref{sec-stub} and \ref{fig:lattices}) and in the Kagomè model (see Section \ref{sec-kago} and \ref{fig:1dkagome}). Indeed, it turns out that in this case, an atom dispersively coupled to the FB seeds a BS with features analogous to typical BSs close to the band edge of an isolated {\it dispersive} band [\eg as in \ref{fig:saw} (c)-(e)]. This is witnessed by \ref{fig:1dkagome}(c) for the Kagomè model, which shows that the BS localization length gets larger and larger as the detuning from the FB decreases in contrast to the saturation behavior next to an isolated FB that occurs \eg in \ref{fig:saw} (f)-(h). Notice that in these 1D examples the dispersive band scales quadratically in the vicinity of the FB.

\begin{figure}[t]
    \centering\includegraphics[width=0.6\textwidth]{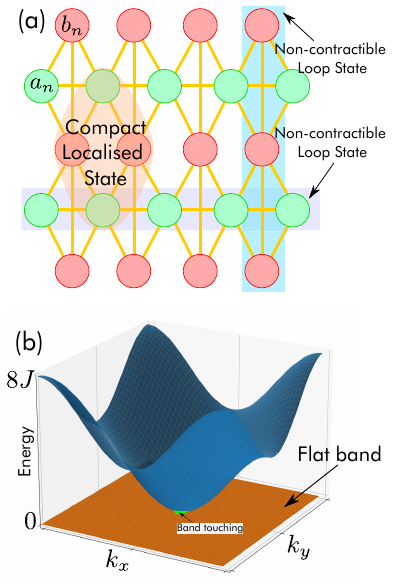}
    \caption{(a) Structure of the  checkerboard lattice and the set of CLSs. For completeness, we also display the pair of non-contractible loop states which must be added to the CLSs to form a complete basis of the FB subspace (see footnote in the main text). (b) Spectrum of the 2D checkerboard model, showing a band touching (green dot) between the dispersive band (blue) and the FB (orange) at zero energy. }
	\label{fig:checkerboard}
\end{figure}

FBs falling on the edge of dispersive bands can also occur in 2D. A paradigmatic instance is the checkerboard lattice in \ref{fig:checkerboard}(a). This a bipartite lattice where nearest-neighbour cavities of the same sublattice are coupled with hopping rate $-J$, while the hopping rate between nearest-neighbour cavities of different sublattices is $-J$ along one diagonal and $J$ along the other diagonal. Each cavity has frequency $2J$. 
The spectrum features a FB and a dispersive band according to [see also \ref{fig:checkerboard}(b)]
\begin{align}\label{checker}
	\omega_{\rm FB} = 0, &&
	\omega_{d}(k_x,k_y) = 2J\qty[2-\cos(k_x)-\cos(k_y)].
\end{align} 
The two bands touch one another at the $\Gamma$ point $k_x=k_y=0$ (close to this $\omega_{d}(k_x,k_y)$ has a parabolic shape). A set of non-orthogonal CLSs, one for each unit cell, is easily identified as [see \ref{fig:checkerboard}(b)]
\begin{equation}
	\label{eq:CLScheck}
	\ket{\phi_{n}} = \tfrac{1}{2} \Big[\ket{a_{n_x,n_y}}{-}\ket{a_{n_x-1,n_y}} {+} \ket{b_{n_x,n_y}} {-} \ket{b_{n_x,n_y+1}}\Big]\,,
\end{equation}
where $n=(n_x,n_y)$ labels unit cells \footnote{Unlike previous examples of overlapping CLSs, the set \eqref{eq:CLScheck} is not a complete basis of the FB eigenspace, which can be shown to be due to the singular behaviour of the present FB \cite{rhim_classification_2019}. To get a complete basis, the set \eqref{eq:CLScheck} must be complemented with a pair of so called non-contractible loop states [see \ref{fig:checkerboard}(b)]. These two states are yet unbound, hence in the thermodynamic limit we can neglect their contribution to the BS (this arising from the local atom-field interaction).}. 
We are thus in the case $U=(2,2)$ with the overlap between nearest-neighbour CLSs being given by $\alpha_x = \alpha_y = \alpha = 1/4$. 

Now, remarkably, we numerically checked that an atom dispersively coupled to the FB at $\omega_{\rm FB} = 0$ gives rise to a BS whose features accurately match those predicted in Section \ref{sec:2d} for $\alpha=1/4$, meaning that localization length is independent of the detuning and has the finite value  $\lambda_{\rm BS}\simeq 0.567$ [recall the discussion after \eq\eqref{eq:lbs2d}]. In other words, we find that the behaviour is identical to that of an energetically isolated FB. 

The above case studies can be understood by recalling that a dispersive band that scales quadratically in the vicinity of a band edge gives a contribution to the density of states (DOS) which is finite in 2D while it diverges in 1D (in the latter case a van Hove singularity occurs). In contrast, the DOS of a FB clearly has $\delta$-like diverging at $\omega_{\rm FB}$. It follows that in 2D the dispersive band can be neglected compared to the FB, explaining why in this case the properties of the BS match those predicted for an isolated FB in Section \ref{sec:2d}. In 1D, instead, the contribution of the dispersive band can no longer be neglected, explaining why \eg in the Kagomè lattice the BS behaviour is different from the one in Section \ref{subsec:loclength}.

\section{Giant atoms coupled to a photonic flat band} 
\label{sec:giants}

In recent years, it has become experimentally possible fabricating so called {\it giant atoms} \cite{frisk_kockum_quantum_2021}, \ie quantum emitters which are coupled non-locally to the field through a discrete set of coupling points (where a standard atom is retrieved in the special case of only one coupling point). 
For a giant atom with ${\cal N}$ coupling points the total Hamiltonian \eqref{eq:tot_lm} is generalized as
\begin{equation}
	\label{eq:tot_lm-2}
	H = \omega_{0} \sigma^\dagger  \sigma +  H_B + \sum_{\ell=1}^{\cal N} \left(g_\ell a^\dagger_{x_\ell}\sigma + \hc \right)\,,
\end{equation}
where $x_\ell$ (coupling point) labels the cavity to which the giant atom is coupled to and $g_\ell$ the corresponding (generally complex) coupling strength.
It is convenient to define the field ladder operator \cite{leonforte_quantum_2024}
\begin{equation}
	a_\chi=\sum_{\ell=1}^{\cal N} \gamma_\ell^* a_{x_\ell}\,\,\,{\rm with}\,\,\,\gamma_\ell=\frac{g_\ell}{\bar{g}}\,\,{\rm and}\,\,\bar{g}=\sqrt{\sum_\ell |g_\ell|^2}\,,\label{achi}
\end{equation}
where, due to $\sum_{\ell=1}^{\cal N} |\gamma_\ell|^2=1$, $a_\chi$ fulfills $[a_\chi,a_\chi^\dag]=1$. With this definition, \eqref{eq:tot_lm-2} can be arranged {\it formally} as the Hamiltonian in the presence of a normal atom
\begin{equation}
	H=H_B+\omega_0 \sigma^\dagger \sigma+\bar{g} \left(a_{\chi}^\dag \sigma +\hc\right)\label{H1-bis}\,
\end{equation}
(notice however that $a_{\chi}$ generally does not commute with field operators $a_{x}^\dag$, this being a signature of the non-local nature of atom-photon coupling). Let us define the single-photon state 
\begin{equation}
\ket{\chi} = a_\chi^\dagger\ket{\rm vac}\,, \label{site-s}
\end{equation}
which for convenience in the remainder we will call \textit{site state} since it can be formally thought to be associated with a fictitious location of the atom.

\subsection{Atom-photon bound state and $H_{\rm eff}$}

It can be straightforwardly shown \cite{leonforte_quantum_2024} that the theory of atom-photon BSs and effective Hamiltonians, which we reviewed in Sections \ref{bs-sec} and \ref{sec-Heff} is naturally generalized to the case of giant atoms once the atom location (previously called $\ket{x_0}$ with only one atom) is replaced with the site state \eqref{site-s}. In particular, the BS of one atom in the dispersive regime now reads $\ket{\psi_{\rm BS}}=G_B(\omega_0) \ket{\chi}$[\cf\eqs\eqref{psibs-1} and \eqref {psibs}].

Interestingly, in the usual regime where the atom is dispersively coupled to an isolated FB, the possibility to engineer the giant atom's coupling points and relative strengths allows for the BS to have just the same shape as a CLS. Indeed, if $\ket{\phi_n}$ is a CLS, then clearly  $G_B(\omega_{0}) \ket{\phi_n}=\ket{\phi_n}/(\omega_0{-}\omega_{\rm FB})$ (since $H_B \ket{\phi_n}=\omega_{{\rm FB}}\ket{\phi_n}$). Accordingly, if we choose the emitter's coupling points such that
\begin{equation}\label{chi-cls}
\ket{\chi}=\ket{\phi_n}
\end{equation}
(for some given $n$) then [see \eq\eqref{psibs} for $\ket{x_0}\rightarrow \ket{\chi}$]
\begin{equation}\label{psi-gia}
\ket{\psi_{{\rm BS}}}=\frac{g}{\omega_0-\omega_{\rm FB}}\ket{\phi_n}\,.
\end{equation}
This shows that the BS has just the same wavefunction as a CLS of the FB. In the case of many giant atoms labeled by $n$ and each such that $\ket{\chi_n}=\ket{\phi_n}$, the photon-mediated interaction strength is then given by [see \eq\eqref{eq:kij} for $\ket{x_i}\rightarrow \ket{\chi_i}$]
\begin{equation}
    \mathcal{K}_{nn'} = \frac{g^2}{\omega_0-\omega_{\rm FB}}\braket{\chi_n}{\chi_{n'}}\,.
\end{equation}
An interesting consequence of this is that for nearest-neighbour CLSs [\eg in the sawtooth model of \ref{fig:saw}(a) or the stub lattice of \ref{fig:lattices}(c)] an effective spin Hamiltonian arises with strictly nearest-neighbour interactions. In the stub lattice, in particular, we get [\cf\eq\eqref{eq:alphadelta}]
\begin{equation}
    \mathcal{K}_{nn'} = \frac{g^2}{\omega_0-\omega_{\rm FB}}\,\frac{1}{2+\Delta}\,,
\end{equation}
hence the interaction strength can be in principle modulated through parameter $\Delta$ (provided that this remains large enough to neglect the effect of the dispersive bands).

Taking advantage of the compact nature of CLSs, this type of dipole-dipole interactions can be implemented in 1D in a relatively straightforward fashion by using giant atoms with very few coupling points (only three in the sawtooth and stub lattices). 

\subsection{General case}

The above is immediately generalized to the case that the site state is an arbitrary superposition of CLSs, \ie when \eq\eqref{chi-cls} is replaced by
\begin{equation}
    \ket{\chi}=\sum_n c_n \ket{\phi_n}\,,
\end{equation}
which through an analogous argument leads to [see \eq\eqref{psi-gia}]
\begin{equation}\label{psi-gia-2}
\ket{\psi_{{\rm BS}}}=\frac{g}{\omega_0-\omega_{\rm FB}}\sum_n c_n \ket{\phi_n}\,.
\end{equation}

\section{Conclusions}
\label{sec-conc}

In this work, we carried out a general study of dipole-dipole interactions between quantum emitters dispersively coupled to photonic flat bands (FBs), \ie bands with flat dispersion law. 

In line with standard theory of dipole-dipole dispersive interactions, the spatial shape of such interactions, hence the interaction range, is inherited from the wavefunction of the atom-photon bound state (BS) arising when the emitter is off-resonant with the band. 
In the case of a FB, such localization length can saturate to a finite value in a way that the BS extends even beyond the cell where the emitter sits, thus enabling dipole-dipole interactions between atoms coupled to different cells. We showed on a general basis that this type of atom-photon BS can be connected with so called compact localized states (CLSs), a key concept in the theory of FBs. 
Remarkably, we showed that for 1D lattices with nearest-neighbour overlapping CLSs the localization length (hence the dipole-dipole interaction range) monotonically grows with the overlap between CLSs, according to a universal law which we derived explicitly, until diverging when the overlap tends to its maximum value. This suggests that the cross-talk between emitters placed in different cells is enabled by the CLSs overlap. An analogous task was carried out in 2D square lattices, which showed that, unlike 1D, the BS localization now converges to a finite value as the CLSs overlap approaches its maximum value. We also considered the singular situation that a FB, instead of being energetically isolated, touches a dispersive band. In such case, we showed that the BS behaves like in the presence of typical dispersive bands in 1D (localization length diverging with the detuning) while in 2D it behaves like in the presence of an isolated FB. Finally, we considered the effect of replacing the emitter with a so called giant atom which can couple non-locally to the lattice to a manifold of distinct cavities, showing that in this case one can engineer the coupling points in a way that the BS wavefunction has just the same shape as a CLS or a linear combination of a few of these.

Unavoidable occurrence of disorder in real setups could affect some of the phenomena predicted here, which is briefly discussed in \aref{app:disorder} in some case studies. Yet, it is reasonable to expect that recent advancements in platforms such as cold atoms \cite{gonzalez-tudela_2019_engineering} and circuit QED \cite{jouanny_band_2024} allow for devices that are clean enough to see FB-induced effects, as recently demonstrated in \rref \cite{chase-mayoral_compact_2024}.

Within the topical framework of atom-photon interactions in unconventional photonic environments, which takes advantage of current experimental capabilities enabling the fabrication of emitters coupled to engineered baths, our work introduces a new paradigm of dipole-dipole interactions. Additionally, it establishes a new link with the hot research area in condensed matter and photonics investigating flat bands \cite{vicencio_poblete_photonic_2021,leykam_artificial_2018}. 

\section*{Acknowledgements} 
\label{sec:acknowledgements}

We are grateful to D. De Bernardis, L. Leonforte, P. Scarlino, V. Jouanny and L. Peyruchat for fruitful discussions and A. Miragliotta for reading the manuscript. E.D.B. acknowledges support from the Erasmus project during his stay at IFF-CSIC. E.D.B. and F.C. acknowledge financial support from European Union-Next Generation EU through projects: Eurostart 2022 ‘Topological atom-photon interactions for quantum technologies’; PRIN 2022–PNRR No. P202253RLY ‘Harnessing topological phases for quantum technologies’; THENCE–Partenariato Esteso NQSTI–PE00000023–Spoke 2 ‘Taming and harnessing decoherence in complex networks’. A.G.T. acknowledges support from the CSIC Research Platform on Quantum Technologies PTI-001 and from Spanish projects PID2021-127968NB-I00 funded by MICIU/AEI/10.13039/501100011033/ and by FEDER Una manera de hacer Europa, and TED2021-130552B-C22 funded by  MICIU/AEI /10.13039/501100011033 and by the European Union NextGenerationEU/ PRTR, respectively, and QUANTERA project MOLAR with reference PCI2024-153449 and funded by MICIU/AEI/10.13039/501100011033 and the European Union.

\appendix

\section{More analytical details on 1D models with flat bands}
\label{app:lattices}

In this appendix, we provide more details on 1D models exhibiting FBs discussed in the main text. Due their 1D nature, for simplicity we call simply $n$ the Bravais lattices identifying the unit cells. Based on the Bloch theorem, one first define traslationally-invariant ladder operators as
\begin{equation}\label{alphak}
    \alpha_{k,\nu} = \frac{1}{\sqrt{N}} \sum_{n} e^{-ik n}\, a_{n,\nu}
\end{equation}
with $\nu=1,...,Q$ labeling the sublattices and $k$ running over the first Brillouin zone. In terms of these, the bare bath Hamiltonian can be arranged in the form
\begin{equation}
    H_B = \sum_{k} \left( \alpha^\dagger_{k,1} \quad \dots \quad \alpha^\dagger_{k, Q} \right) {H}_{k} 
	\left( 
	\begin{array}{c}
		\alpha_{k,1}\\
		\vdots\\
		\alpha_{k,Q}
	\end{array} \right),
\end{equation}
where $H_k$ is a $Q\times Q$ Hermitian matrix (Bloch Hamiltonian) representing the Hamiltonian in the momentum space. Diagonalizing $H_k$ yields the dispersion laws of all bands (indexed by $m$) and the corresponding normal modes.

\subsection{Sawtooth lattice}
\label{subsec:sawtooth}

The Bloch Hamiltonian is the $2\times2$ matrix 
\begin{equation}
    \label{eq:hamsawtooth}
    H_k = J
    \begin{pmatrix}
        0 & \sqrt{2} (1+e^{-ik})\\
        \sqrt{2} (1+e^{ik}) & -2 \cos{(k)}
    \end{pmatrix},
\end{equation}
which is readily diagonalised, resulting in the two dispersion laws in \eq~\eqref{saw-w} and in single-photon eigenstates (each describing a photon populating a normal mode) 
\begin{align}
    \label{eq:eig}
    \ket{\Psi_{{\rm FB}, k}} &\propto \left[(1+e^{-ik}) \ket{a_k} - \sqrt{2}\ket{b_k}\right],\\
    \ket{\Psi_{d, k}} &\propto  \left[\left(\frac{1+e^{-ik}}{1+\cos{(k)}}\right) \ket{a_k} + \sqrt{2}\ket{b_k}\right]
\end{align}
(we omit normalization factors as intended by symbol $\propto$). Here, $\ket{a_k}$ and $\ket{b_k}$ are single-photon states corresponding to \eqref{alphak}, where $a$ and $b$ are the sublattice indexes defined in \ref{fig:saw}(a).
In particular, notice that each $\ket{\Psi_{\rm FB}, k}$ form an orthogonal basis of unbound states spanning the FB eigenspace, which is alternative to the CLS basis of Eq.~\eqref{eq:CLSsaw}.

\subsection{Double-comb lattice}
\label{subsec:double-comb}

The Bloch Hamiltonian in this case reads
\begin{equation}
    H_{k} =
    \begin{pmatrix}
        \omega_c & 0 & t\\
        0 & \omega_c & t\\
        t & t & 2J\cos k
    \end{pmatrix}\,,
\end{equation}
which is easily diagonalized. The eigenvalues embody the dispersion laws of the three bands and read
\begin{equation*}
    \begin{split}
        \omega_{ \rm FB} &= \omega_c,\\
        \omega_{\pm} (k) &=	 \frac{\omega_c}{2} - J \cos k \pm \sqrt{\frac{t^2}{2} + \qty(\frac{\omega_c}{2} + J\cos k)^2}\,.
    \end{split}
\end{equation*}

\subsection{Stub lattice}
\label{subsec:stub}

For the three-partite stub lattice [Hamiltonian parameters and sublattice indexes are defined in \ref{fig:lattices}(c)], the Bloch Hamiltonian can be cast in the form
\begin{equation}
    H_{k} = J
    \begin{pmatrix}
        0 & \sqrt{\Delta} & 0\\
        \sqrt{\Delta} & 0 & 1 + e^{-ik}\\
        0 & 1 + e^{ik} & 0
    \end{pmatrix}
\end{equation}
which, upon diagonalisation, yields the band dispersion laws
\begin{align}
    \omega_{ \rm FB} = 0, && \omega_{\pm} (k) = \pm J \sqrt{\Delta + 2(1+\cos{k})}\,,
\end{align}
so that $\Delta$ in fact measures the gap between the zero-energy FB and either dispersive band. In particular, the FB eigenstates are worked out as
\begin{equation}
    \ket{\Psi_{ {\rm FB},k}} \propto \left[(1 + e^{ik}) \ket{a_k} - \sqrt{\Delta} \ket{c_k}\right].
\end{equation}
and form an orthogonal basis of unbound states alternative to the non-orthogonal basis of CLSs \eqref{eq:CLSstub}.

\subsection{1D Kagomè lattice} \label{sec-kago}

This five-partite lattice represents a sort of 1D counterpart of the Kagomè model \cite{maimaiti_universal_2019}, where the Hamiltonian parameters and sublattice indexes are defined in \ref{fig:1dkagome}(c).
The Bloch Hamiltonian $H_k$ is calculated as the $5\times5$ matrix given by
\begin{equation}
    H_k = 
    \begin{pmatrix}
        0 & 1-e^{-ik} & -e^{-ik} & 0 & 0\\
        1-e^{ik} & 0 & -1 & 0 & 0\\
        -e^{ik} & -1 & 0 & -1 & -e^{ik}\\
        0 & 0 & -1 & 0 & 1-e^{ik}\\
        0 & 0 & -e^{-ik} & 1-e^{-ik} & 0
    \end{pmatrix}\,,\!\!
\end{equation}
giving rise to five bands, one of which is a FB of frequency $\omega_{\rm FB} = 2J$ touching with the upper edge of a dispersive band. 

\section{FB projector in the CLS basis} \label{app:proj}

To derive \eq\eqref{eq:CLSPFB} we need to express projector \eqref{eq:PSIPFB} in the CLSs basis which requires the expansion of each FB Bloch eigenstate in the CLS basis
\begin{equation}
    \ket{\Psi_{\vb{k},{\rm FB}}} = \frac{1}{N^{D/2}} \sum_n a_{\vb{k},n} \ket{\phi_n}\,.\label{psikfb}
\end{equation}
To determine the expansion coefficients $a_{\vb{k},n}$, we express \eq\eqref{psikfb} in matrix-vector form as 
\begin{equation}
    \vb{v}_{\vb{k}} = {\bf M} \cdot \vb{a}_{\vb{k}}\,,  \label{lin-prob}
\end{equation}
where $a_{\vb{k},n}$ is the $N$-dimensional column vector having coefficients $a_{\vb{k},n}$ as components while $\vb{v}_{\vb{k}}$ and ${\bf M}$ are respectively the $N$-dimensional column vector and $N\times N$ matrix having entries [we use \eq\eqref{eq:cls-expansion}]
\begin{equation}
    v_{\vb{k},n} = \braket{\phi_n}{\Psi_{\vb{k},{\rm FB}}} = \sqrt{f(\vb{k})}\, e^{i\vb{k}\cdot \vb{r}_n}\,,\,\,
    M_{nn'} = \braket{\phi_n}{\phi_{n'}}.
\end{equation}
Now, owing to translation symmetry, the overlap matrix $M_{nn'}$ (which is positive definite) can be written as
\begin{equation}
    \label{app:eq:Mij}
    \begin{split}
    M_{nn'} &= \delta_{nn'} + \sum_{d=1}^D \alpha_{1,d} (\delta_{n,n'+\vu{e}_d} + \delta_{n,n'-\vu{e}_d}) + \\
    &+ \sum_{d=1}^D \alpha_{2,d} (\delta_{n,n'+2\vu{e}_d} + \delta_{n,n'-2\vu{e}_d}) + \dots,  
    \end{split}
\end{equation}
where we sum over nearest-neighboring cavities, next-to nearest and so on to cavity $n$ along the $d$th direction and accordingly $\alpha_{1,d}$ ($\alpha_{2,d}$) is the overlap along said direction.
This choice also specifies the shape of function $f$, according to the equality between \Eqs \eqref{eq:braket} and \eqref{app:eq:Mij}, which holds if 
\begin{equation}
  f(\vb{k}) = 1 + 2\sum_{d} \alpha_{1,d}\cos(k_d) + 2\sum_{d} \alpha_{2,d}\cos(2k_d)+\dots\,,\label{fkapp} 
\end{equation}
where $k_d$ denotes the $d$th component of the $\vb{k}$ vector. We also see that these are basically the eigenvalues of $M$ (due to translation symmetry).

Inverting \eq\eqref{lin-prob} we get $a_{\vb{k}} = \textbf{M}^{-1}\cdot\vb{v}_{\vb{k}}$ and exploiting again translation symmetry we obtain 
\begin{equation}
    \ket{\Psi_{\vb{k},{\rm FB}}} = \frac{1}{N^{D/2}} \frac{1}{\sqrt{f(\vb{k})}} \sum_n e^{i\vb{k}\cdot \vb{r}_n} \ket{\phi_n}\,.
\end{equation}
By plugging this into \eq\eqref{eq:PSIPFB} we finally end up with \eq\eqref{eq:CLSPFB}.

\begin{figure}
    \centering
    \includegraphics[width=\textwidth]{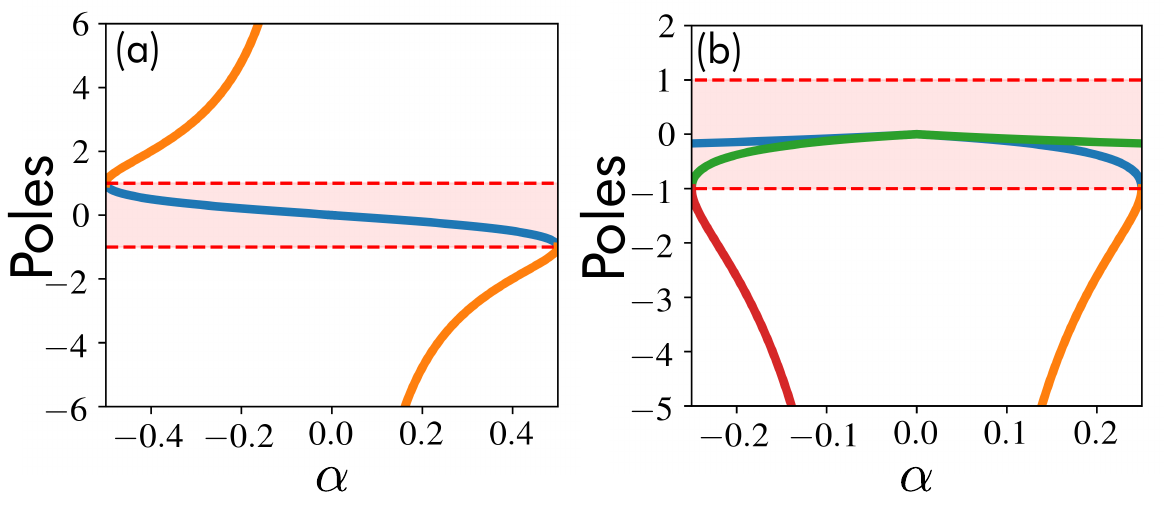}
    \caption{Poles of the integrand of $\xi_{nn'}$ in the 1D and 2D cases [respectively panels (a) and (b)]  plotted as a function of the CLS overlap $\alpha$ (different colours correspond to different poles). When one of plotted functions takes values within the region highlighted in red, the corresponding pole falls within the the unit circle, \ie inside the integration contour of the complex integral. In (b), notice that for $|\alpha| = 1/4$, one of the two poles stays well inside the unit circle, corresponding to a BS with finite localization length.}
    \label{fig:poles}
\end{figure}

\section{Explicit form of function $\xi_{nn'}$}\label{app:xi}

Here, we show the derivation of the explicit form of function $\xi_{nn'}$ which was given in the main text.

\subsection{1D case}

For 1D lattices and CLSs of class $U=2$ (only nearest-neighbour CLSs overlap), the scalar product between any pair of CLSs has the form $\braket{\phi_n}{\phi_{n'}} = \delta_{n,n'} + \alpha\, \delta_{n,n'+ 1} + \alpha\, \delta_{n,n'- 1}$. Hence, function $f(k)$ takes the form [\cf\eq\eqref{fk}]
\begin{equation}
    f(k) = 1 + 2 \alpha\,\cos k,
\end{equation} 
with $|\alpha|\leq1/2$. Replacing this in \eq (28) for $D=1$ yields
\begin{equation}
    \xi_{nn'} = \frac{1}{N} \sum_k \frac{e^{ik(n-n')}}{1+2\alpha\cos k},
\end{equation}
where, since we are working in 1D, we made the replacement $x_n = n$. Based on periodic boundary conditions, $k$ takes $N$ values within the first Brillouin zone $\comm{-\pi}{\pi}$ with uniform spacing $\Delta k = 2\pi/N$. Accordingly, we can rewrite $\xi_{nn'}$ as
\begin{equation}
    \xi_{nn'} = \frac{1}{2\pi} \sum_k \Delta k \frac{e^{ik(n-n')}}{1+2\alpha\cos k}\,.
\end{equation}
In the thermodynamic limit $N\to\infty$, $k$ becomes a continuous variable and the sum is turned into a continuous integral
\begin{equation}
    \label{app:eq:1d}
    \xi_{nn'} = \frac{1}{2\pi} \int_{-\pi}^{\pi} \dd k\,  \frac{ e^{ik\,|n-n'|} }{1 + 2 \alpha\,\cos k}\,. 
\end{equation}

\noindent 
Notice that in the exponent we replaced $n-n'$ with its modulus since the integral is invariant under the substitution $n-n' \to n'-n$. 

Applying a standard technique \cite{economou_greens_1979,arfken_chapter_2013},
\eqref{app:eq:1d} can be expressed as integral on the unit circle on the complex plane upon the change of variable $z = e^{ik}$ (since $-\pi< k \leq \pi$, $z$ lies on the unit circle of the complex plane). Accordingly, $\dd z = i e^{ik} \dd k = iz\dd k$ and $\cos{k} = (e^{ik}+e^{-ik})/2 = (z+z^{-1})/2 = (z^2+1)/(2z)$.
This yields
\begin{equation}
    \xi_{nm} = \frac{1}{2\pi i} \oint \dd z  \frac{ z^{|n-n'|} }{\alpha z^2 + z + \alpha},
\end{equation}
This expression can be computed with the help of the residue theorem by noting that, since $|\alpha|<1/2$, the integrand function always has a pole falling inside the unit circle at $z_0 = -\frac{1}{2\alpha} \left(1-\sqrt{1-4\alpha^2}\right)$ [see blue curve in \ref{fig:poles}(a)]. 
Calculating the corresponding residue we thus end up with
\begin{equation}
    \xi_{nn'} = \frac{(-\sgn\alpha)^{|n-n'|}}{\sqrt{1-4\alpha^2}} \, e^{-\frac{|n-n'|}{\lambda_{\rm 1D}}}\,,
\end{equation}
with the localization length $\lambda_{\rm 1D}$ given by \eq~\eqref{eq:xi1d}.

\subsection{2D case}

In 2D and for a square geometry, \eqref{fkapp} reduces to
\begin{equation}
    f(k_x,k_y) = 1+2\alpha_x\cos(k_x) + 2\alpha_y\cos(k_y),
\end{equation}
with $\alpha_x=\alpha_{1,1}$, $\alpha_y=\alpha_{1,2}$ (according to notation introduced in \Eq~\eqref{app:eq:Mij}) and subject to the constraint $|\alpha_x|+|\alpha_y|< 1/2$.
Plugging into \eq~\eqref{xinn} and taking the thermodynamic limit as done previously in the 1D case produces the double integral
\begin{equation}
    \xi_{nn'} = \frac{1}{(2\pi)^2} \iint_{\rm BZ} \dd k_x \dd k_y \frac{e^{i k_x (n-n')} e^{i k_y (m-m')}}{1+2\alpha_x\cos(k_x) + 2\alpha_y\cos(k_y)}\,,
\end{equation}
where \textit{BZ} stands for the region of integration specified by $-\pi\le k_x\le \pi$ and $-\pi\le k_y\le \pi$. 
To evaluate the integral, we first take $m=m'$ (\ie we look at the behaviour of $\xi_{nn'}$ along the $\vu{e}_x$ direction) which allows to write the integral as
\begin{align}
    \xi_{nn'} =& \frac{1}{2\pi} \int_{-\pi}^{\pi} \dd k_x e^{ik_x(n-n')} \nonumber\\
    &\times \left(\frac{1}{2\pi} \int_{-\pi}^{\pi} \dd k_y \frac{1}{1+2\alpha_x\cos(k_x) + 2\alpha_y\cos(k_y)}\right),
\end{align}
which can be solved by applying the residue theorem twice, first to the integral over $k_y$ and then to that over $k_x$. There occur four poles given by 
\begin{align}
    z_{1,\pm} = \frac{x_+}{2} \pm \frac{\sqrt{x_+^2 - 4}}{2},\,\,\, z_{2,\pm} = \frac{x_-}{2} \pm \frac{\sqrt{x_-^2 - 4}}{2}\,
\end{align}
with
\begin{equation}
x_{\pm} = \frac{-1\pm2\sgn{(\alpha_x)}|\alpha_y|}{|\alpha_y|}\,.
\end{equation}
We henceforth focus on the isotropic case $\alpha_x = \alpha_y = \alpha$ with $|\alpha| \leq 1/4$, which ensures that the localization length calculated along the $x$-direction will be match the one along any other direction. In this isotropic case, the only two poles falling inside the unit circle are given by
\begin{equation}
    \begin{split}
    z_{1,+} &= \frac{2\alpha-1}{2|\alpha|} + \frac{\sqrt{1-4\alpha}}{2|\alpha|}, \\
    z_{2,+} &= -\frac{2\alpha+1}{2|\alpha|} + \frac{\sqrt{1+4\alpha}}{2|\alpha|}\,.
    \end{split}
\end{equation}
Calculating the corresponding residues and summing up as prescribed by the residue theorem we end up with \eq\eqref{xinn2D}. 

\section{BS localization length from $\xi_{nn'}$}
\label{sec-corr}

In this appendix, we show that the BS scales with distance in the same way as function $\xi_{nn'}$ defined in \eq\eqref{xinn}. For the sake of argument, we will focus on CLSs of class $U=2$ in 1D
 (the 2D case for $=(2,2)$ is treated analogously on each of the two directions).

Let $\ket{n,\nu}$ the state where a single photon lies on the $n$th unit cell in the $\nu$th sublattice. Since we assume $U=2$, the CLS is localized only on two nearest-neighbor cells, say $n$ and $n+1$, so that its wavefunction can be written as
\begin{equation}\label{CLS-2}
    \ket{\phi_n} = \sum_{\nu} c_{1,\nu}\ket{n,\nu} + c_{2,\nu}\ket{n-1,\nu}\,,
\end{equation}
where $c_{i,\nu}$ are the (in general complex) coefficients of the CLS in front of state $\ket{n,\nu}$, which do not depend on $n$ due to translational invariance.  
If the atom is coupled to cavity $(n_0,\nu_0)$, the weight function $w_n(x_0)$ \Eq \eqref{weight} will take thus the form
\begin{equation}
    w_n(x_0) = c_{0,\nu_0}^* \xi_{nn_0} + c_{1,\nu_0}^* \xi_{nn_0+1}.
\end{equation}
Thus, owing to \Eq \eqref{psibs-3}, the shape of the bound state will be given by
\begin{equation}
    \begin{split}
        \ket{\psi_{\rm BS}} &\propto \sum_{n\nu} \Big[ c_{0,\nu}(c_{0,\nu_0}^* \xi_{nn_0} + c_{1,\nu_0}^* \xi_{nn_0+1})\ket{n,\nu}\\
        &+ c_{1,\nu}(c_{0,\nu_0}^* \xi_{nn_0} + c_{1,\nu_0}^* \xi_{nn_0+1})\ket{n-1,\nu}\Big]\,,
    \end{split}
\end{equation}
which, upon shifting $n-1\to n$ in the second term can be turned into
\begin{equation}
    \begin{split}
    \ket{\psi_{\rm BS}} &\propto \sum_{n\nu} \Big[(c_{0,\nu}c_{0,\nu_0}^*+c_{1,\nu}c_{1,\nu_0}^*)\xi_{nn_0}\\
    &+c_{0,\nu}c_{1,\nu_0}^*\xi_{nn_0+1}+c_{1,\nu}c_{0,\nu_0}^*\xi_{nn_0-1}\Big]\ket{n,\nu}\\
    &=\sum_{n\nu} A_{n\nu} \ket{n,\nu}\,.
    \end{split}
\end{equation}
Now, if $\xi_{nn_0} \propto \exp{-\frac{|n-n_0|}{\lambda}}$ (exponential scaling), evidently coefficient $A_{n\nu}$ will also have the same exponential scaling with $n$. 
This shows that in general that the BS wavefunction scales exponentially with the same localization length as function $\xi_{nn'}$. An analogous result holds if $\xi_{nn'}$ is the sum of a finite number of exponentially localised terms.

The proof can be extended to CLSs of arbitrary class $U$ as long as the weight function $\xi_{nn'}$ scales as the sum of a finite number of exponentials. Notice that in the case of a complex pole $z$ in the computation of $\xi_{nn'}$, the associated localization length will be given by $\lambda^{-1} = \log |z|$.

\begin{figure}[h]
   \centering
   \includegraphics[width=\linewidth]{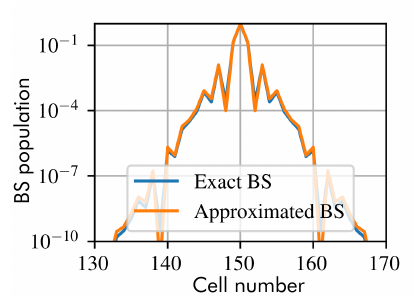}
   \caption{Population of the atom-photon bound state on sublattice $a$ of the lattice in Fig. 3(d) of \rref \cite{maimaiti_universal_2019} (the behavior is similar on other sublattices). There is a good agreement between the BS as obtained from exact diagonalization (blue line) and the one predicted by our theory (orange line).}
    \label{fig:U3}
\end{figure}

\section{Flatbands of class $U=3$}
\label{app:u3}

In the main text, we on FBs where only nearest-neighbor CLSs overlap, \ie $U=2$ for 1D lattices, due to their experimental feasibility in state-of-the-art experimental platforms. Nonetheless, our formalism can be naturally extended also to more general cases, such as $U>2$ in 1D.

To provide evidence of this, in this Appendix we show how out general theory can be extended to one-dimensional lattices of class $U=3$. In this case, function $f(k)$ is written as $f(k)=1+2\alpha_{1,x}\cos{k}+2\alpha_{2,x}\cos{2k}$, where $\alpha_{1,x} = \braket{\phi_n}{\phi_{n\pm1}}$ and $\alpha_{2,x} = \braket{\phi_n}{\phi_{n\pm2}}$ [see \Eq \eqref{app:eq:Mij} for notation]. Thus, $\xi_{nn'}$ can be written in the thermodynamic limit as (see \aref{app:xi})
\begin{equation}
    \xi_{nn'} = \frac{1}{2\pi} \int_{-\pi}^{\pi} \dd k \frac{e^{ik\abs{n-n'}}}{1+2\alpha_{1,x}\cos{k}+2\alpha_{2,x}\cos{2k}}
\end{equation}
which, upon the substitution $z=e^{ik}$ can be turned into a complex integral on the unit circle reading
\begin{equation}
    \label{app:xiu3}
    \xi_{nn'} = \frac{1}{2\pi i} \oint \dd z \frac{z^{\abs{n-n'}+1}}{\alpha_{2,x} z^4 + \alpha_{1,x} z^3 + z^2 + \alpha_{1,x} z + \alpha_{2,x}}\,.
\end{equation}
Since its denominator is a symmetric quartic polynomial, the poles of the integrand function can be analytically worked out as
\begin{align}
    z_1^{\pm} &= \frac{-\beta_+ \pm \sqrt{2\alpha_{1,x}\beta_+-8\alpha_{2,x}^2-4\alpha_{2,x}}}{4\alpha_{2,x}}\,,\\
    z_2^{\pm} &= \frac{-\beta_- \pm \sqrt{2\alpha_{1,x}\beta_--8\alpha_{2,x}^2-4\alpha_{2,x}}}{4\alpha_{2,x}}\,,
\end{align}
where $\beta_{\pm} = \alpha_{1,x} \pm \sqrt{\alpha_{1,x}^2 + 8\alpha_{2,x}^2 - 4\alpha_{2,x}}$.
These are two pairs of complex and conjugate numbers, meaning that $0$, $2$ or $4$ poles will exists within the unit circle, depending on the overlap parameters $\alpha_{1,x}$ and $\alpha_{2,x}$. This ensures that \eqref{app:xiu3} is always real.

To test the above, we considered the three-partite one-dimensional photonic lattice considered in Fig. 3(d) of \rref \cite{maimaiti_universal_2019}. Such lattice consists of three sublattices, namely $a,b,c$, described by the Hamiltonian
\begin{equation}
    \begin{split}
        &H_B/J = \sum_n \qty(a^\dagger_n b_n + b_n^\dagger c_n + \hc)\\
        & + \sum_n \Big[a_n^\dagger \qty(-0.523\, a_{n+1} + 0.17\, b_{n+1} + 0.693\, c_{n+1})\\
        & + b_n^\dagger \qty(-0.627\, a_{n+1} - 0.115\, b_{n+1} + 0.512\, c_{n+1}) \\
        & + c_n^\dagger \qty(-0.731\, a_{n+1} - 0.399\, b_{n+1} + 0.332\, c_{n+1}) + \hc \Big]
    \end{split}
\end{equation}
where $J$ is an energy constant. In this lattice, a $U=3$ FB appears at energy $\omega_{\rm FB} = -1.5J$, whose CLSs read
\begin{equation}
    \begin{split}
        \ket{\phi_n} &\propto \ket{a_n} - \ket{b_n} + \ket{c_n}+ 0.255 \ket{a_{n+1}}\\
        &+ 0.286 \ket{b_{n+1}}- 0.599 \ket{c_{n+1}} + 0.25 \ket{a_{n+2}}\\
        &- 0.5 \ket{b_{n+2}} + 0.25 \ket{c_{n+2}}\,.
    \end{split}
\end{equation}
By coupling an atom dispersively to this FB, we checked that the atom-photon BS agrees with our theory, which is shown in \ref{fig:U3}. Interestingly, such BSs has a non-monotonic spatial shape.


\section{Effect of disorder}
\label{app:disorder}


Inevitable occurrence of disorder in real experimental setups can affect FBs \cite{chau_2024_disorder-induced, martinez_2023_flat-band-localization}, hence it is natural to ask how the predicted effects in this work are affected by disorder in the photonic lattice. While a comprehensive study of the effects of disorder is beyond the scope of this work, in this Appendix we discuss briefly some case studies in 1D in the presence of disorder on the cavity frequencies (diagonal disorder) and on the photon hopping rates (off-diagonal disorder). For diagonal disorder, we change the photonic bath Hamiltonian as $H_B \to H_B + \sum_{x} \omega_x(\eta) a^\dagger_x a_x$, while for off-diagonal disorder as $H_B \to H_B + \sum_{x\not =x'} J_{xx'}(\eta) (a^\dagger_{x'}a_x+\hc)$. Here, $\omega_x(\eta)$ and $J_{xx'}(\eta)$ are random values drawn from a uniform probability distribution in the interval $\comm{-\eta}{\eta}$ (in units of $J$).
We will address separately FBs with orthogonal and non-orthogonal Compact Localised States (CLSs) in the case studies of the double-comb lattice and stub lattice, respectively.

In line with the  manuscript, as a case study of flatbands with {\it orthogonal CLSs} we consider the {\it double-comb lattice} (see \ref{fig:lattices}a-b), whose relevant parameters (without disorder) are the cavity frequencies $\omega_c$ and hopping rates $J$ and $t$ (see also \aref{subsec:double-comb}). 
We focus on the representative case $J = 1, \omega_c = 4J, t = 3J$, $\omega_0 = 3.8J, g = 0.01J$. With this choice of parameters, a FB arises with energy $\omega_{\rm FB} = \omega_c$, which is well-separated from the dispersive bands (allowing us to discern effects due to disorder on the FB from the standard Anderson localization in the dispersive bands). Diagonal disorder is simulated by introducing random cavity frequencies for sublattice $b$, while off-diagonal disorder concerns the two hopping rates $J$ and $t$.
In the case of an off-diagonal disorder in $J$, due to the peculiar shape of the CLSs in this system (whose amplitudes vanish on the $b$-sublattice), neither the FB nor the shape of the CLSs is affected by adding disorder on $J$. As such, dipole-dipole interactions mediated by the FB are unchanged. On the other hand, in the case of off-diagonal disorder on $t$, by calling $t_{bc}^n$ ($t_{ac}^n$) the local hopping rate between cavities $b_n$ ($a_n$) and $c_n$, it is immediately seen (through the usual destructive interference argument) that any state $\ket{\phi_n} \propto t_{bc}^n \ket{a_n} - t_{ac}^n \ket{c_n}$ is a CLS with energy $\omega_{\rm FB}=\omega_c$. Therefore, despite the disorder, the FB still exists at the same frequency where each CLS now features slightly different amplitudes on the two sublattices.         
Accordingly, short-ranged photon-mediated interactions between emitters still arise in this case with an interaction strength that will not depend on the local values of $t_{bc}^n$ and $t_{ac}^n$.
Finally, in the case of diagonal disorder, as expected, the flatness of the FB is spoiled, which however does not cause a qualitative significant change of the predicted effects. In particular, we find that if $\eta \lesssim g$ the atom-atom interaction strength is preserved, while if $\eta > g$ the interaction strength can be enhanced or lowered, depending on specific realization of disorder. The interaction range is preserved in both cases.         
This striking behaviour somehow is not consistent with the localization-delocalization transition in FBs \cite{chau_2024_disorder-induced} due to disorder, i.e., the fact that FB eigenstates are delocalized with weak disorder and localized with strong disorder (standard Anderson localization is recovered in the latter case). 
We conjecture that no significant effects arise from this transition as long as there exists a relatively large energy gap between the FB and the dispersive bands (as is the case here given the considered choice of parameters). 
Arguably, an enhanced interaction range might arise with weak disorder if the gap were smaller, which however might affect the validity of the single FB approximation (holding in the regime $g \ll \delta_{\rm FB} \ll \delta_d$).
    
We next address {\it non-orthogonal CLSs} in the case study of the {\it stub lattice} [see \ref{fig:lattices}(c-(d)], which is a tripartite lattice where cavities $b_n$ and $c_n$ are coupled to each other with hopping equal to $J$ and cavities $b_n$ are side-coupled to cavities $a_n$ with strength $J\sqrt{\Delta}$ ($\Delta$ being a dimensionless parameter. 
Recall that the energy gap between the FB and each dispersive band is measured by $J\sqrt{\Delta}$. Again, we study the effect of diagonal and off-diagonal disorder. We recall that in this lattice the existence of a zero-energy FB is guaranteed by chiral symmetry [see \aref{subsec:stub}]. For our simulations, we set $J = 1, \Delta = 1.5$, $\omega_0 = 0.1J$ and $g = 0.01J$. 
In the case of off-diagonal disorder on $J$, where we introduce disorder on both the $a_n$-$b_n$ and $b_n$-$c_n$ hopping rates, chiral symmetry is not broken, hence the zero-energy FB is unaffected. However, the shape of CLSs (which are non-orthogonal) is changed by the disorder and so are the resulting atom-photon bound states and mediated interactions. As is reasonable to expect, such changes are weak for $\eta \lesssim g$, hence the localization length and interaction strength are unaffected in this regime. Instead, in the regime $\eta \gtrsim J\sqrt{\Delta}$, the  localization length gets shorter and the interaction strength weaker, which can be attributed to the standard effect of Anderson localization. In the intermediate regime, i.e., when $g<\eta<J\sqrt{\Delta}$ we observe signatures of the localization-delocalization transition, entailing an increase in the localization length of the bound states due to delocalization of the CLSs, which results in the increase of the interaction strength between emitters coupled to distant unit cells.
On the other hand, introducing diagonal disorder on all cavities, chiral symmetry no longer holds and the FB is no longer flat. Again, we observe a similar behavior as in the former case (i.e., when $g<\eta<J\sqrt{\Delta}$). Indeed, in this regime the interaction strength between distant emitters might be enhanced for certain disorder realizations. This can be seen as a signature of the delocalization of the CLSs due to weak disorder and, as a consequence, an increase of the localization length of the bound state.

The above discussion provides evidence that there exist regimes where the essential phenomena predicted in our work can show up even in the presence of disorder. 
A comprehensive study of the effect of disorder goes beyond the scope of the manuscript and it will be the subject of a future work.

\bibliography{DBGTC_v1}

\clearpage

\end{document}